\documentclass[aps,prd,reprint,preprintnumbers,superscriptaddress,nofootinbib,longbibliography,floatfix]{revtex4-2}
\pdfoutput=1

\usepackage[utf8]{inputenc} 
\usepackage[T1]{fontenc}    
\usepackage{hyperref}       
\usepackage{url}            
\usepackage{booktabs}       
\usepackage{amsfonts}       
\usepackage{nicefrac}       
\usepackage{microtype}      
\usepackage{xcolor}         
\usepackage{amsmath}
\usepackage{graphicx,subcaption}
\usepackage{float}
\usepackage{amsmath}
\usepackage{physics}
\usepackage{xcolor}
\usepackage{multirow}
\usepackage{comment}
\usepackage{float}

\hypersetup{
  colorlinks=true,
  citecolor=blue,
  linkcolor=blue,
  urlcolor=blue
}

\newcommand{\fff}{{Flows for Flows}\xspace}

\begin{document}

\title{Flows for Flows: Morphing one Dataset into another \\ with Maximum Likelihood Estimation}

\author{Tobias Golling}
\email{tobias.golling@unige.ch}
\affiliation{Département de Physique Nucléaire et Corpusculaire, Université de Genève, Genève; Switzerland}

\author{Samuel Klein}
\email{samuel.klein@unige.ch}
\affiliation{Département de Physique Nucléaire et Corpusculaire, Université de Genève, Genève; Switzerland}

\author{Radha Mastandrea}
\email{rmastand@berkeley.edu}
\affiliation{Department of Physics, University of California, Berkeley, CA 94720, USA}
\affiliation{Physics Division, Lawrence Berkeley National Laboratory, Berkeley, CA 94720, USA}

\author{Benjamin Nachman}
\email{bpnachman@lbl.gov}
\affiliation{Physics Division, Lawrence Berkeley National Laboratory, Berkeley, CA 94720, USA}
\affiliation{Berkeley Institute for Data Science, University of California, Berkeley, CA 94720, USA}

\author{John Andrew Raine}
\email{john.raine@unige.ch}
\affiliation{Département de Physique Nucléaire et Corpusculaire, Université de Genève, Genève; Switzerland}

    \begin{abstract}
          
          Many components of data analysis in high energy physics and beyond require morphing one dataset into another.  This is commonly solved via reweighting, but there are many advantages of preserving weights and shifting the data points instead. Normalizing flows are machine learning models with impressive precision on a variety of particle physics tasks.  Naively, normalizing flows cannot be used for morphing because they require knowledge of the probability density of the starting dataset.  In most cases in particle physics, we can generate more examples, but we do not know densities explicitly.  We propose a protocol called \textbf{flows for flows} for training normalizing flows to morph one dataset into another even if the underlying probability density of neither dataset is known explicitly.  This enables a morphing strategy trained with maximum likelihood estimation, a setup that has been shown to be highly effective in related tasks.  We study variations on this protocol to explore how far the data points are moved to statistically match the two datasets.  Furthermore, we show how to condition the learned flows on particular features in order to create a morphing function for every value of the conditioning feature.  For illustration, we demonstrate flows for flows for toy examples as well as a collider physics example involving dijet events.
          \vfill
    \end{abstract}

\maketitle

    \section{Introduction}



One common data analysis task in high energy physics and beyond is to take a reference set of examples $R$ and modify them to be statistically identical to a target set of examples $T$.  In this setting, we do not have access to the probability density of $x\in\mathbb{R}^N$ responsible for $R$ or $T$ (i.e. $p_T$ and $p_R$), but we can sample from both by running an experiment or simulator. Examples of this task include shifting simulation to match data for detector calibrations, morphing experimental or simulated calibration data to match backgrounds in signal-sensitive regions of phase space for background estimation or anomaly detection, and tweaking simulated examples with one set of parameters to match another set for parameter inference.

A well-studied way to achieve dataset morphing is to assign importance weights $w$ so that $w(x)\approx p_T(x)/p_R(x)$.  This likelihood ratio can be constructed using machine learning-based classifiers (see e.g.~\cite{hastie01statisticallearning,sugiyama_suzuki_kanamori_2012}) to readily accommodate $N\gg 1$ without ever needing to estimate $p_T$ or $p_R$ directly.  While highly effective, likelihood-ratio methods also have a number of fundamental challenges.  With non-unity weights, the statistical power of a dataset is diluted.  Furthermore, even small regions of non-overlapping support between $p_T$ and $p_R$ can cause estimation strategies for $w$ to fail.  

A complementary strategy to importance weights is direct feature morphing.  In this case, the goal is to find a map $f:\mathbb{R}^N\rightarrow\mathbb{R}^N$ from the reference to the target space such that the probability density of $f(x\sim p_R)$ matches $p_T$. Unlike the importance sampling scenario, $f$ is not unique.  The goal of this paper is to study how to construct $f$ as a \textit{normalizing flow}~\cite{tabak_flows,flows_review} -- a type of invertible deep neural network most often used for density estimation or sample generation.  Normalizing flows have proven to be highly effective generative models, which motivates their use as morphing functions.  Traditionally, normalizing flows are trained in the setting where $p_R$ is known explicitly (e.g. a Gaussian distribution). Here we explore how to use flows when neither $p_R$ or $p_T$ are known explicitly. We call our method \textbf{flows for flows}.  This approach naturally allows for the morphing to be conditional on some feature, such as a mass variable~\cite{Raine:2022hht,Golling:2022nkl,Sengupta:2023xqy}.  Approaches similar to flows for flows have been performed for variational autoencoders~\cite{Howard:2021pos} and, recently, diffusion models~\cite{Diefenbacher:2023flw}.


In many cases in physics, $p_R$ is close to $p_T$, and so $f$ should not be far from the identity map. For example, $R$ might be a simulation of data $T$, or $R$ might be close to $T$ in phase space.  In order to assess how well suited normalizing flows are for this case, we also study how much $x$ is moved via the morphing. An effective morphing map need not move the features minimally, but models that include this inductive bias may be more robust than those that do not.  There is also a connection with optimal transport, which would be exciting to study in the future.


This paper is organized as follows.  Section~\ref{sec:methods} briefly reviews normalizing flows and introduces all of the flows for flows variations we study.  Next, Sec.~\ref{sec:results_toy} presents a simple application of the flows for flows variations on two-dimensional synthetic datasets. Sec.~\ref{sec:science} gives a more realistic application of the transport variations to sets of simulated particle collision data. We summarize the results and conclude in Sec.~\ref{sec:conclusions}.

    \section{Methods}
\label{sec:methods}

\subsection{Normalizing flows as transfer functions}
\label{sec:nf_transfer}

Normalizing flows are classically defined by a parameteric diffeomorphism $f_\phi$ and a base density $p_\theta$ for which the density is known.
Using the change of variables formula, the log likelihood (paramaterized by both $\theta$ and $\phi$) of a data point $x \sim p_D$ under a normalizing flow is given by
\begin{equation}
    \log p_{\theta, \phi} (x) = \log p_\theta (f_\phi^{-1}(x)) - \log \left| \det (J_{f_\phi^{-1}(x)}) \right|,
\end{equation}
where $J$ is the Jacobian of $f_\phi$.
Training the model to maximise the likelihood of data samples results in a map $f_\phi^{-1}$ between the data distribution $p_D(x)$ and the base density $p_\theta$. As the base density should have a known distribution, it is usually taken to be a normal distribution of the same dimensionalty as the data (which motivates the name ``normalizing'' flow). 

At this point, we can introduce the first transfer method from a reference distribution $p_R$ to a target distribution $p_T$, the \textbf{base transfer}. For this method, we train two normalizing flows with two different maps from the same base density. If $f_{\phi_1}$ constitutes a map to the reference density $p_R$ and $f_{\phi_2}$ is a map to the target density $p_T$, then the composition $f_{\phi_2}^{-1} \circ f_{\phi_1}$ is a transfer map $f:R\rightarrow T$. In other words, the transfer method routes from reference to target via some base density intermediary. 

It is also possible to use a learned base density, such as another normalizing flow, instead of some known base distribution. This is our second method, \textbf{unidirectional transfer}. Given samples from two data distributions $p_{R}$ and $p_{T}$ of the same dimensionality, a map $f_{\gamma}:R\rightarrow T$ between these distributions can be found by estimating a density $p_{\phi,R}$ for $R$ to use as the base density in the construction of another normalizing flow. 
In practice, this involves first training a normalizing flow to learn the density $p_R$ by constructing the map $f_{\phi}^{-1}$ from a base density $p_\theta$ to $p_R$.

Training of the two normalizing flows (the first for the base density, the second for the transport) is done by maximising the log likelihood of the data under the densities defined by the change of variables formula and given by 
\begin{align*}
    \begin{split}
    &\max_\gamma \mathop{\mathbb{E}}_{y \sim p_{T}} \left[ \log p_{\theta, \phi, \gamma}(y) \right]
    \\
    &= \max_\gamma \mathop{\mathbb{E}}_{y \sim p_{T}} \left[ \log p_{\theta, \phi}(f_\gamma^{-1}(y)) - \log \left| \det (J_{f_\gamma^{-1}(y)}) \right| \right]; 
    \\
    &\max_\phi \mathop{\mathbb{E}}_{x \sim p_{R}} \left[ \log p_{\theta, \phi}(x) \right]
    \\
    &= \max_\phi \mathop{\mathbb{E}}_{x \sim p_{R}} \left[ \log p_{\theta}(f_\phi^{-1}(x)) - \log \left| \det (J_{f_\phi^{-1}(x)}) \right| \right].
    \end{split}
\end{align*}

\begin{figure}
    \centering
    \includegraphics[width=0.45\textwidth]{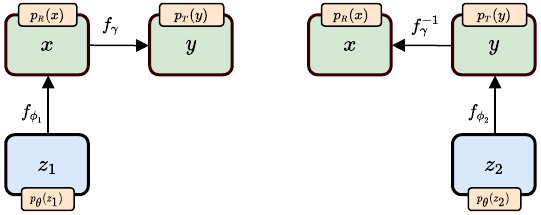}
    \caption{A schematic of the flows for flows architecture.}
    \label{fig:fff_schematic}
\end{figure}

As a direct extension of the unidirectional training method: defining densities on both the reference and the target distributions, $p_{\theta_1, R}$ and $p_{\theta_2, T}$ allows both $f_\gamma$ and $f_\gamma^{-1}$ to be explicitly used by training in both directions, from $R$  to $T$ and $T$ to $R$. This comprises our third transfer method, \textbf{flows for flows}.
A benefit of training in both directions is that the dependence of $f_\gamma$ on the defined and learned densities $p_{\theta_1, R}$ and $p_{\theta_2, T}$ is reduced.
A schematic of the flows for flows architecture is shown in Fig.~\ref{fig:fff_schematic}.

The invertible network $f_\gamma$ that is used to map between the two distributions may not have semantic meaning on its own as some invertible neural networks are known to be universal function approximators. 
This map can become interpretable if it is subject to additional constraints. In this work, we investigate two physically-motivated modifications to the flow training procedure: \textbf{movement penalty}, where we add an L1 loss term to the flow training loss, and \textbf{identity initialization}, where we initialize the flow architecture to the identity function.  The L1 variation directly penalizes the average absolute value of the distance moved, while the idea for the identity initialization is that the model will converge on the first best solution that gives close to no movement. All five transfer methods introduce in this section are summarized in Tab.~\ref{tab:transfer_methods}.

\begin{table}[]
    \centering
    \begin{tabular}{|c|c|}
        \hline
        Method name & Training heuristic  \\
        \hline
        \hline
        Base transfer & $p_R \rightarrow \mathcal{N}(0,1) \rightarrow p_T$ \\
        \hline
        Unidirectional transfer & $p_R \rightarrow p_T$  \\
        \hline
        Flows for flows & $p_R \longleftrightarrow p_T$ \\
        \hline
        Movement penalty & $p_R \stackrel{+L1}{\longleftrightarrow} p_T$ \\
        \hline
        Identity initialization & $p_R (\mathbb{I} + \epsilon\leftrightarrow) p_T$ \\
        \hline
    \end{tabular}
    \caption{We consider five transfer methods from a reference dataset to a target dataset, both with unknown distributions $p_R$ and $p_T$.}
    \label{tab:transfer_methods}
\end{table}

This entire setup can be made conditional by making the parameters of the invertible neural network dependent on some selected parameter (i.e. the ``condition''). The log-likelihood for a normalizing flow conditioned on some variables $c$ is defined by
\begin{equation}
    \log p_{\theta, \phi} (x | c) = \log p_\theta (f_{\phi(c)}^{-1}(x) | c) - \log \left| \det (J_{f_{\phi(c)}^{-1}(x)}) \right|,
\end{equation}
where the base density can also be conditionally dependent on $c$.
In the case of conditional distributions with continuous conditions, the distributions on data $p_D(x | c)$ will often change smoothly as a function of the condition. For these situations, a flow that is explicitly parameterized by a well-motivated choice of conditioning variable may have a cleaner physical interpretation. We provide an example of such a flow for our application to particle collision datasets in Sec.~\ref{sec:science}. In particular, conditional flows have been used often in high energy physics to develop ``bump hunt'' algorithms to search for new particles \cite{Nachman:2020lpy,Stein:2020rou,hallin2021classifying,Raine:2022hht,Golling:2022nkl,Sengupta:2023xqy}. In such studies, the resulting flows perform well when interpolated to values of the conditioning variable not used in training.

\begin{figure}
    \centering
    \includegraphics[width=0.45\textwidth]{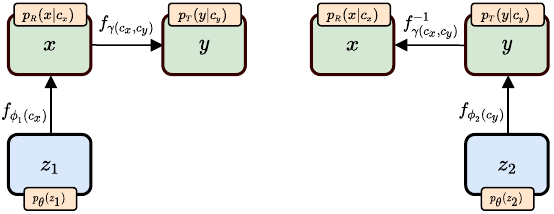}
    \caption{Schematic of a conditional flows for flows architecture.}
    \label{fig:schematic_flow4flow_conditional} 
\end{figure}

A schematic of a conditional flows for flows model is shown in Fig.~\ref*{fig:schematic_flow4flow_conditional}, where the conditioning function $f_{\gamma(c_x, c_y)}$ can also take more restrictive forms, such as $f_{\gamma(c_x - c_y)}$ to ensure that the learned map is simple~\cite{Raine:2022hht,Sengupta:2023xqy}.
Furthermore, the two conditional base distributions can be identical such that $\phi_1 = \phi_2$.
Alternatively the base distributions can be different and instead a shared condition can be use $c=c_x=c_y$.



\subsection{Network architecture}
\label{sec:network_architecture}

Throughout this work, we use two different flow architectures, one for the ``standard'' normalizing flow architecture (i.e. learning transformations from standard normal distributions to arbitrary distributions) and one for the flows for flows architecture (i.e. learning transformations between two nontrivial distributions). 

For the former architecture type, the invertible neural networks are constructed from rational quadratic splines with four autoregressive (AR) layers~\cite{durkan2019neural}. 
Each spline transformation has eight bins and the parameters of the spline are defined using masked AR networks with two blocks and $128$ nodes as defined in the \texttt{nflows} package~\cite{nflows}. For the latter architecture type, we use eight AR layers with splines of eight bins from 3 masked AR blocks of 128 nodes. This slightly more complex architecture is found to give better performance for the large shifts between the toy distributions that we consider. However, in cases where the reference and the target distributions are similar to each other, the architecture of the flows for flows model could in principle be simplified for faster training time while maintaining good performance.

An initial learning rate of $10^{-4}$ is annealed to zero following a cosine schedule~\cite{cosine_annealing} over 60 epochs for the first flow type and 64 epochs for the second flow type. All trainings use a batch size of $128$ and the norm of the gradients is clipped to five. For the toy distribution analyses in Sec.~\ref{sec:results_toy}, the training datasets all contain $10^6$ samples.

    \section{Toy Example Results}
\label{sec:results_toy}

In this section, we explore the performance of the five transfer methods for learning a mapping between nontrivial two-dimensional distributions. In general, we consider both the \textit{accuracy} of the transform -- i.e. does the transfer method learn to successfully morph between the reference and the target distribution -- and the \textit{efficiency} of the transform -- i.e. does the method learn a morphing that is logical, and not unnecessarily circuitous.

\subsection{Base transfer vs. flows for flows}

\begin{figure*}
    \centering
\includegraphics[width = .8\textwidth]{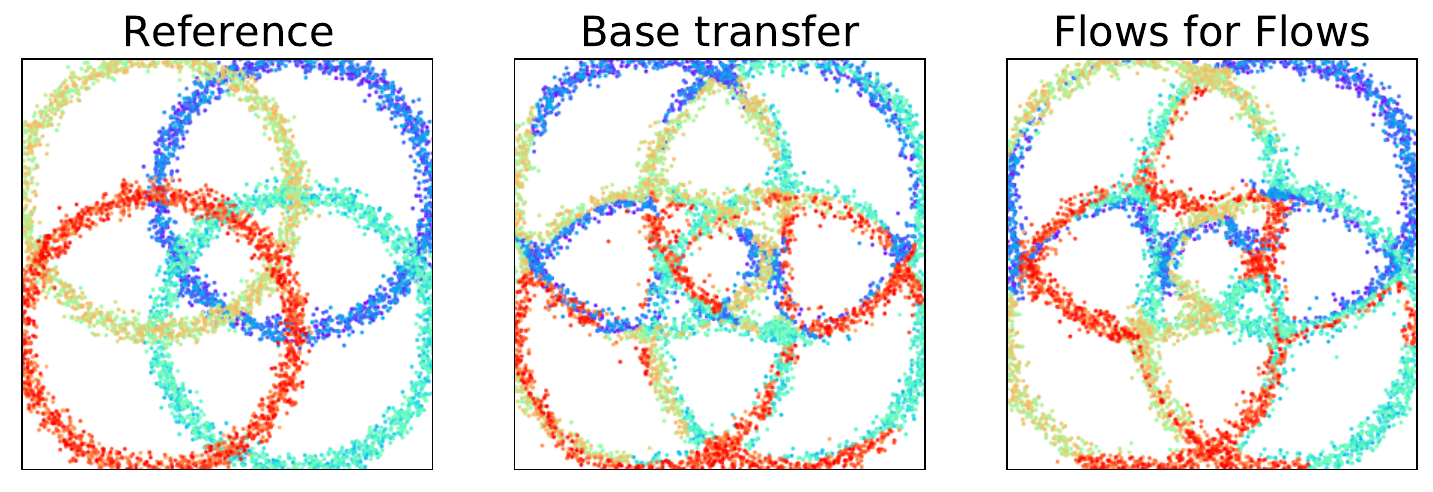}
    \caption{Transport task between two instantiations of the same distribution. The first column shows the reference distribution; the second column shows the base transfer method acting on the reference distribution; the third column shows the flows for flows method. Individual samples have been color coded so as to make clear their paths assigned by the transport method.}
    \label{fig:fourcircles_fourcircles_base_transfer_f4f}
\end{figure*}

In Fig.~\ref{fig:fourcircles_fourcircles_base_transfer_f4f}, we show a transport task between two datasets drawn from a toy distribution of four overlapping circles. Here we are in some sense trying to learn the \textit{identity} mapping. We compare the action of the base transfer, which can be seen as the ``default'' method of mapping between two nontrivial distributions, against the flows for flows method. Both methods are able to successfully map the overall shape of the reference to the target distribution. However, the base transfer method tends not to keep points in the same circle when mapping them from reference to target, while the flows for flows method is more successful at keeping larger portions of each ring together.

\begin{figure*}
    \centering
    \includegraphics[width = .8\textwidth]{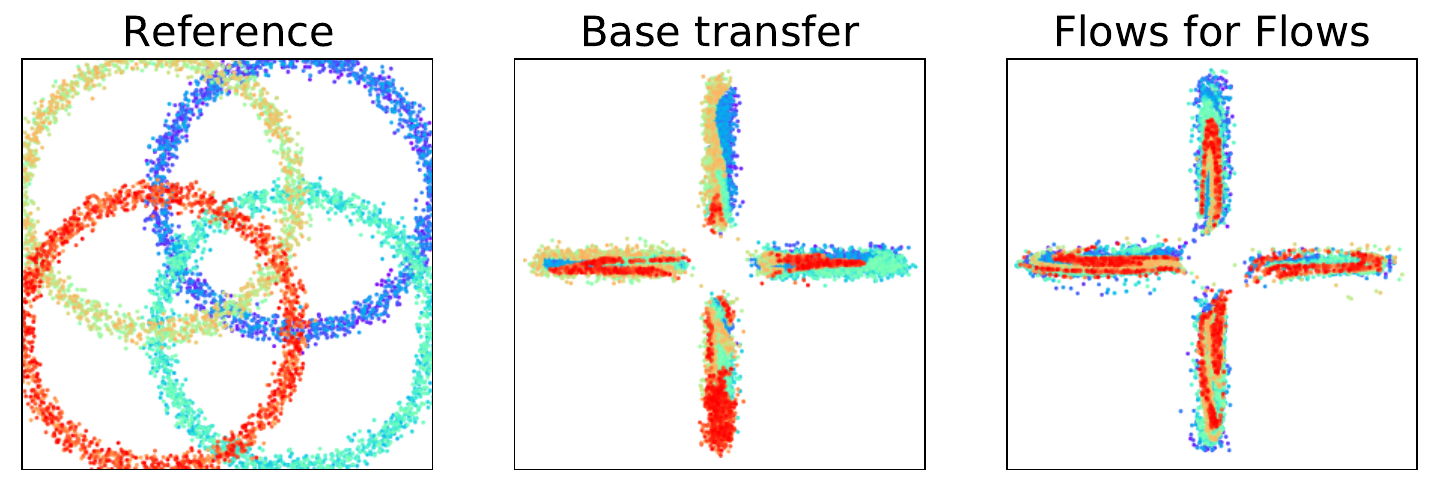}
    \caption{Transport task between two different distributions. Individual samples have been color coded so as to make clear their paths assigned by the transport method. }
    \label{fig:fourcircles_star_base_transfer_f4f}
\end{figure*}

In Fig.~\ref{fig:fourcircles_star_base_transfer_f4f}, we show a transport task between two different distributions, from four overlapping circles to a four-pointed star. As before, both the base transfer and flows for flows method are able to morph the shape of the reference distribution into the shape of the target distribution. Interestingly, the flows for flows method appears to distribute points from each of the four circles more equally among each point of the star.

\subsection{Evaluating multiple transfer methods.}

\begin{figure*}
      \centering
    \includegraphics[width = \textwidth]{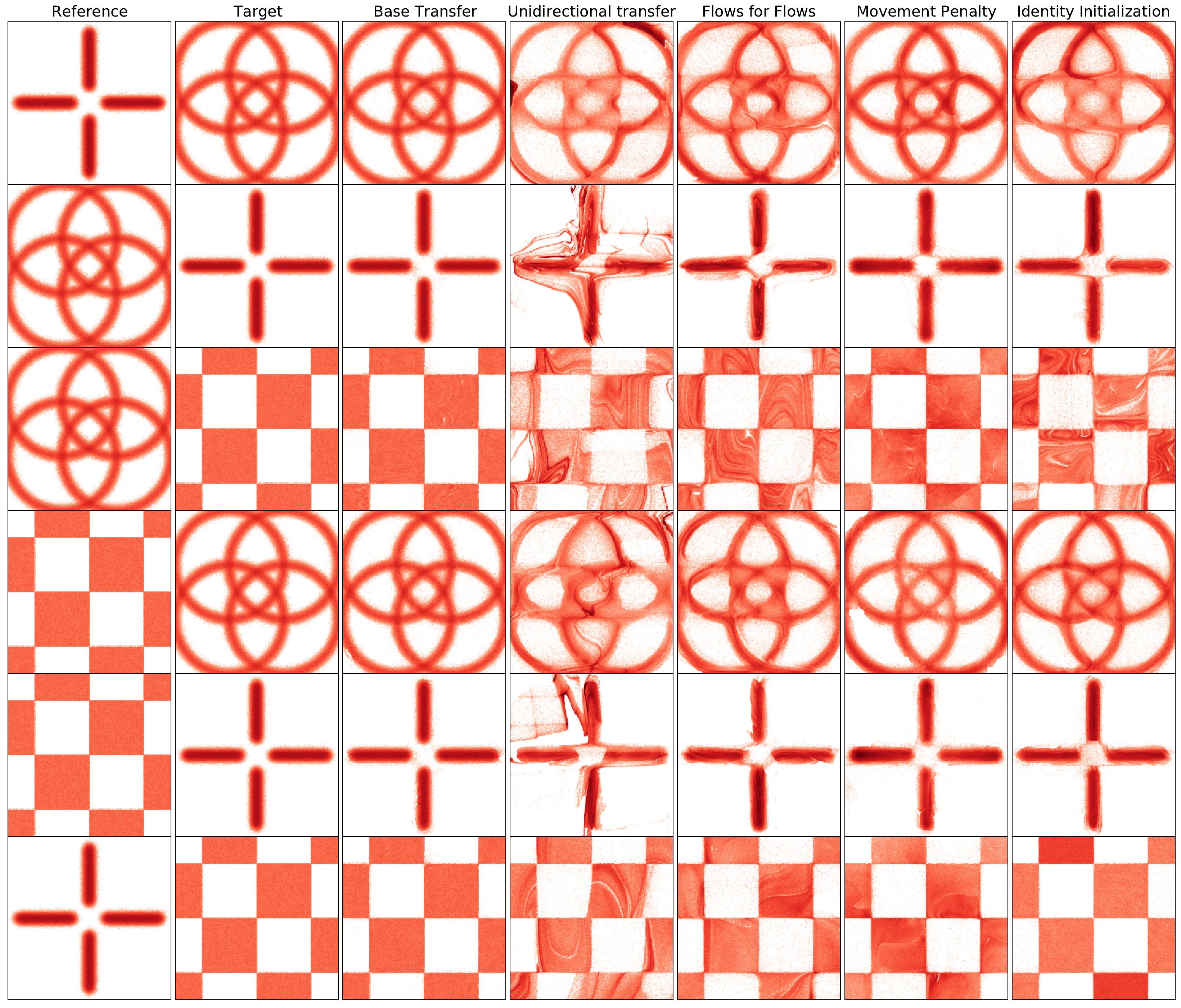}
    \caption{Transport tasks between various choices of nonidentical reference and target toy distributions. The colorbar has been set to scale logarithmically, which can emphasize out-of-distribution points.}
    \label{fig:summary_transfer}
\end{figure*}

In Fig.~\ref{fig:summary_transfer}, we evaluate just the shape-morphing ability of the transfer methods. We consider six reference -- target pairings, where the reference and target distributions are different\footnote{For results corresponding to transports between identical distributions, see App.~\ref{app:plots}.}, and show the action of the base transfer, unidirectional transfer, flows for flows, movement penalty, and identity initialization methods on the reference distribution. We consider transports between three toy distribution types: four overlapping circles, a four-pointed star, and a checkerboard pattern. All of the transfer methods considered are able to successfully learn to map from reference to target, except for the unidirectional transfer, which exhibits a large amount of smearing in the final distribution. Overall, the base transfer, movement penalty, and identity initialization methods show the cleanest final-state distributions.

Another useful metric is the distance traveled by a sample that is mapped under a flow action. For many physical applications, a map that moves data the least is ideal, but we have only explicitly added an L1 loss term to the movement penalty method. Therefore it is interesting to consider how far, on average, all the methods move the features.

\begin{figure}
    \centering
    \includegraphics[width=.4\textwidth]{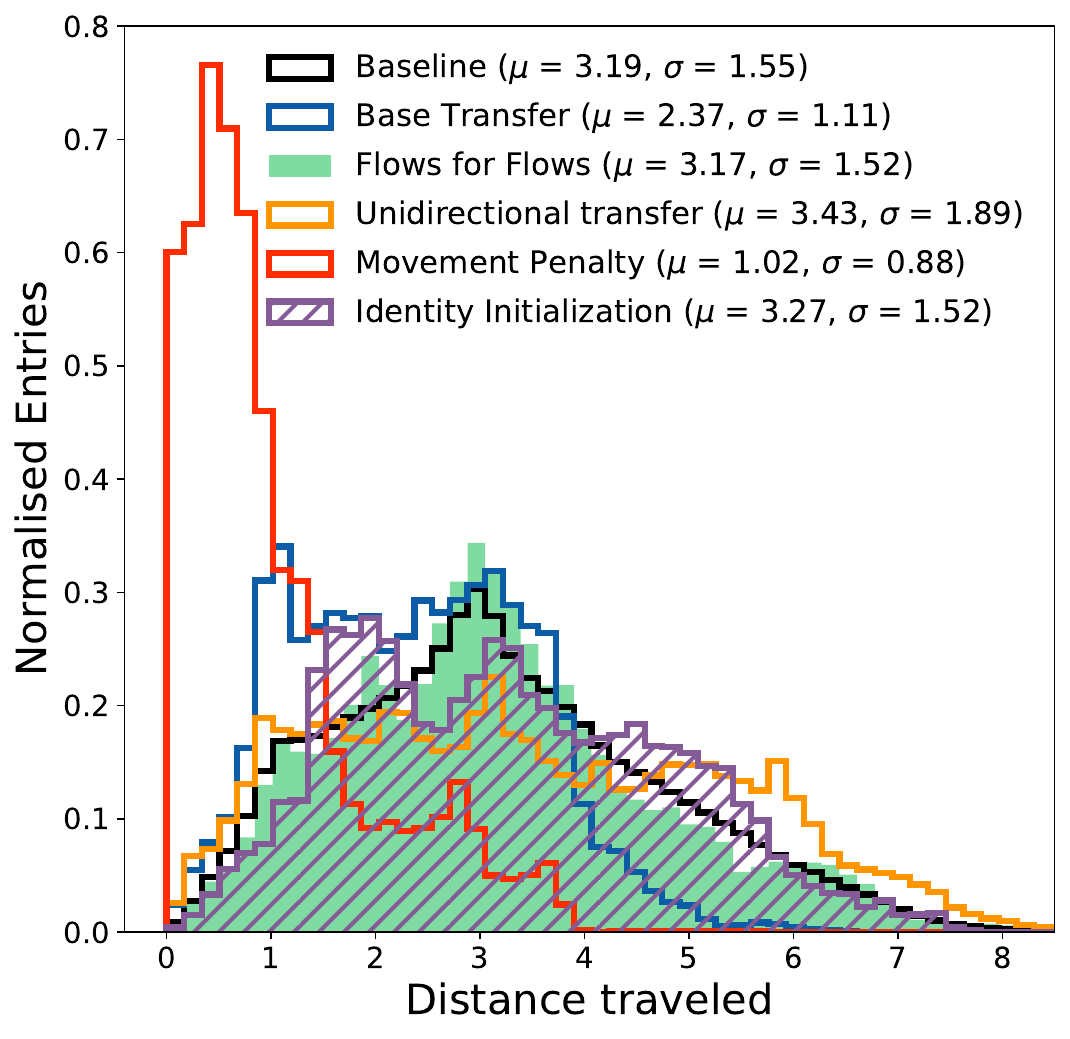}
    \caption{Distances traveled in parameter space between two nonidentical toy distributions. Each histogram compiles data from six transfer tasks, corresponding to the rows of Fig.~\ref{fig:summary_transfer}. The ``baseline'' method shows the distances between two random, unrelated instantiations of each reference -- target distribution pairing. The maximum possible distance travelable in parameter space is 11.31. }
    \label{fig:distances_traveled_transfer}
\end{figure}

In Fig.~\ref{fig:distances_traveled_transfer}, we show a histogram of the distances traveled, amassing all of the six transfer tasks shown in the rows of Fig.~\ref{fig:summary_transfer} so as to equalize over many types of starting and target shapes. The movement penalty method performs best, producing the shortest distances traveled from reference to target by a large margin compared with the other methods. Interestingly, the flows for flows and identity initialization methods have larger mean distances traveled than the base transfer method as well as larger standard deviations. This is somewhat counterintuitive given that the base transfer method does not explicitly link the reference and target distributions during the training procedure, but it may reflect the somewhat contrived nature of the toy examples (especially in light of the more intuitive results for the science datasets in Fig. \ref{fig:science_distances_traveled_total}). All methods except the unidirectional transfer perform more optimally than or on par with the expected baseline, which comes from computing the distances between two random, unrelated instantiations of each reference -- target distribution pairing.

    \section{Application: calibrating collider datasets}
\label{sec:science}

We now move to a physical example: mapping between distributions of scientific observables. Many analyses of collider data are geared towards finding evidence of new physics processes. One powerful search strategy is to compare a set of detected data with an \textit{auxiliary} dataset, where the auxiliary dataset is known to contains Standard Model-only physics. Any nontrivial difference between the detected and the auxiliary datasets could then be taken as evidence for the existence of new physical phenomena. 

The above analysis procedure is contingent upon the auxiliary dataset being a high-fidelity representation of Standard Model physics. However, such an assumption is not true for many datasets that would be, at first glance, ideal candidates for the auxiliary dataset, such as simulation of Standard Model processes or detected data from adjacent regions of phase space. Therefore it is necessary to \textit{calibrate} the auxiliary dataset such that it becomes ideal. Historically, this calibration task has been performed using importance weights estimated from ratios of histograms, either using data-driven approaches like the control region method or fully data-based alternatives.  Recently, machine learning has enabled these approaches to be extended to the case of many dimensions and/or no binning -- see e.g. Ref.~\cite{Karagiorgi:2021ngt} for a review. 

With the flows for flows method, we can consider yet another calibration approach: to create an ideal auxiliary dataset (the target) by morphing the features from a less-ideal, imperfect auxiliary dataset (the reference). When the imperfect auxiliary dataset is chosen to be close to the ideal reference dataset, as would be true of the candidates listed in the previous paragraph, then the flows for flows method should simply be a perturbation on the identity map.%
\footnote{This procedure is the underlying motivation for the Flow-Enhanced Transportation for Anomaly Detection method \cite{Golling:2022nkl}. Equally the ideal and reference could be defined using the same dataset but selecting different conditional values. This method of flows for flows is used to train the Constructing Unobserved Regions with Maximum Likelihood Estimation method~\cite{Sengupta:2023xqy} and is studied for toy datasets in App.~\ref{app:conditional_dists}.}

\subsection{Analysis procedure and dataset}

We focus on the problem of \textit{resonant} anomaly detection, which assumes that given a resonant feature $M$, a potential new particle will have $|M-M_0| \lesssim s$ (which defines the \textit{signal region}) for some unknown $M_0$ and often knowable $s$~\cite{Kasieczka:2021tew}. The value of $M_0$, which corresponds to the mass of the new particle, can be derived from theoretical assumptions on the model of new physics or can be found through a scan. Additional features $X\in\mathbb{R}^N$ are chosen which can be used to distinguish the signal (the new particle) from background (Standard Model-like collisions), which can be done by comparing detected data with reference data within the signal region.

\begin{figure*}[h!]
    \centering
    \includegraphics[width = 1\linewidth]{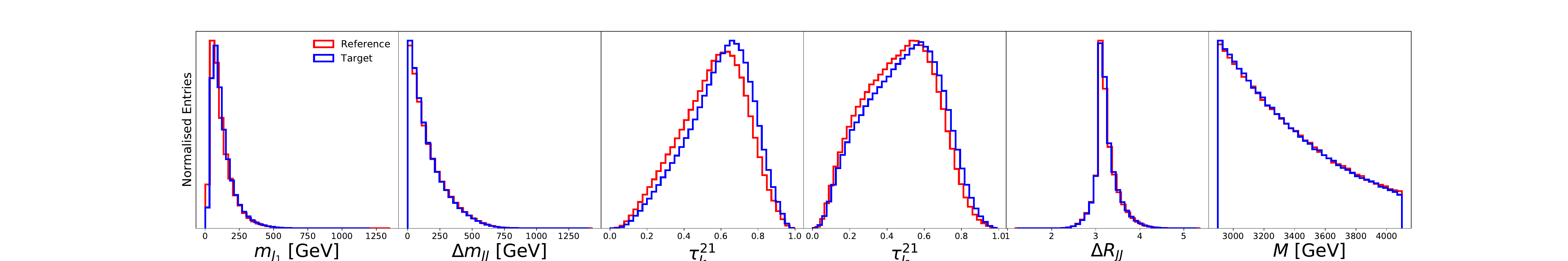}
    \caption{Reference and target distributions used in the application of the flows for flows procedure to scientific datasets. The feature space is comprised of the resonant feature $M$ and five other features $m_{J_1}$, $\Delta m_{JJ}$, $\tau^{21}_{J_1}$, $\tau^{21}_{J_2}$, and $\Delta R_{JJ}$. A description of these observables can be found in \cite{Thaler:2010tr}. The signal region is defined by $|M-M_0|<c$ for $M_0=3.5$ TeV and $c=200$ GeV. }
    \label{fig:features_f4f}
\end{figure*}

For our datasets, we use the LHC 2020 Olympics R\&D dataset~\cite{LHCOlympics,Kasieczka:2021xcg} which consists of a large number ($\sim 10^6$) Standard Model simulation events. The events naturally live in a high-dimensional space, as each contains hundreds of particles with momenta in the $x$, $y$, and $z$ directions. To reduce the dimensionality, the events are clustered into collimated sprays of particles called \textit{jets} using the \textsc{FastJet} \cite{Cacciari:2011ma, Cacciari:2005hq} package with the anti-$k_t$ algorithm \cite{Cacciari:2008gp} ($R$ = 1). From these jets, we can pull a compressed feature space of only five dimensions; this set of features has been extensively studied in collider analyses. The jet features, along with the resonant feature $M$, are displayed in Fig.~\ref{fig:features_f4f}. We take the band $M \in $ [3.3, 3.7] TeV as our signal region.

The LHC Olympics dataset contains two sets of Standard Model data generated from the different simulation toolkits \textsc{Pythia}~8.219~\cite{Sjostrand:2006za,Sjostrand:2014zea} and \textsc{Herwig}++\cite{B_hr_2008}. We use the former as a stand-in for detected collider data. The latter is used as the reference dataset, the less-than-ideal auxiliary dataset that is calibrated through the flows for flows method to form the ideal auxiliary, target dataset. 

To construct the ideal auxiliary dataset, we train a flow to learn the mapping between the reference dataset and the target data \textit{outside} of the signal region, so as to keep the signal region blinded. Once trained, the flow can then be applied to the non-ideal auxiliary dataset within the signal region, thus constructing the ideal auxiliary dataset. We use the same architectures as in Sec. \ref{sec:network_architecture}, with the modification that we \textit{condition} the transport flows on the mass feature $M$. This conditioning is motivated by the fact that the flow is trained outside the signal region and applied within the signal region, which is defined exactly by the variable $M$.

\subsection{Results}

\begin{figure*}
      \centering
    \includegraphics[width = \textwidth]{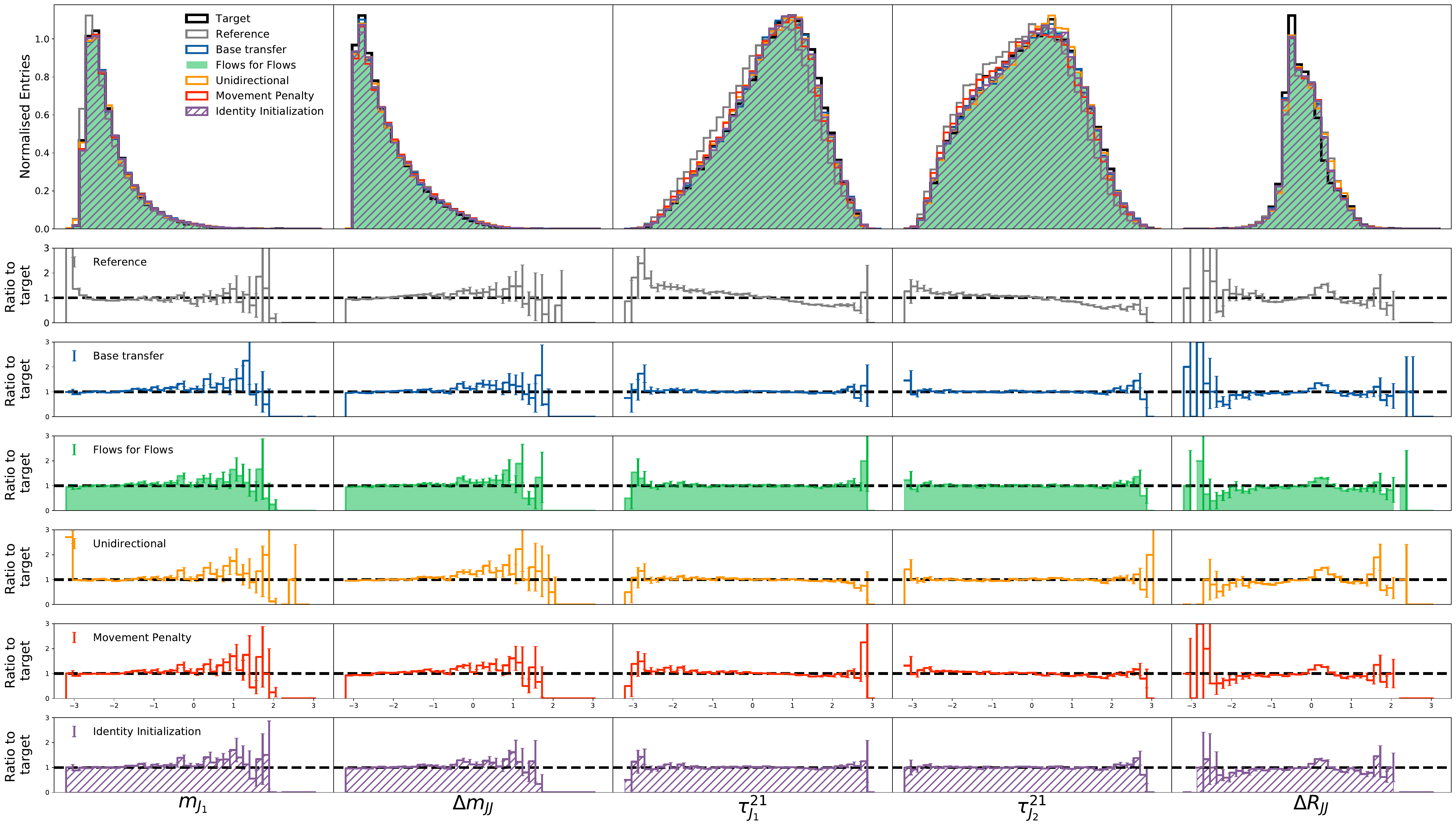}
    \caption{Distributions of and ratios of the flow-transported reference (less-than ideal auxiliary) dataset to the target (ideal auxiliary) dataset. Ratios are taken over each of the five marginal distributions in the parameter space; errorbars represent Poisson uncertainties in bin counts. All data is taken \textit{outside} of the signal region. All features have been individually minmaxscaled to the range [-3, 3] to optimize network training.}
    \label{fig:science_feature_ratios}
    
\end{figure*}

In Fig.~\ref{fig:science_feature_ratios}, we provide the distributions of the flow-moved reference dataset to the target dataset, as well as the ratios to the target, \textit{outside} of the signal region. As is clear from Fig.~\ref{fig:features_f4f}, the reference and target datasets are far more similar in this calibration example than they were in the toy examples. Therefore for the movement penalty method, it was necessary to scan over the strength of the L1 term added to the training loss in order to achieve good performance; we found that we needed to reduce the strength by a factor of 20, as compared with what was used for the toy distributions. In fact, all five transfer methods methods (base transfer, unidirectional transfer, flows for flows, movement penalty, and identity initialization) perform comparably, and all five methods are able to successfully transform the reference dataset such that the five marginal feature distributions greatly resemble those of the target.

\begin{figure}
    \centering
    \includegraphics[width=.4\textwidth]{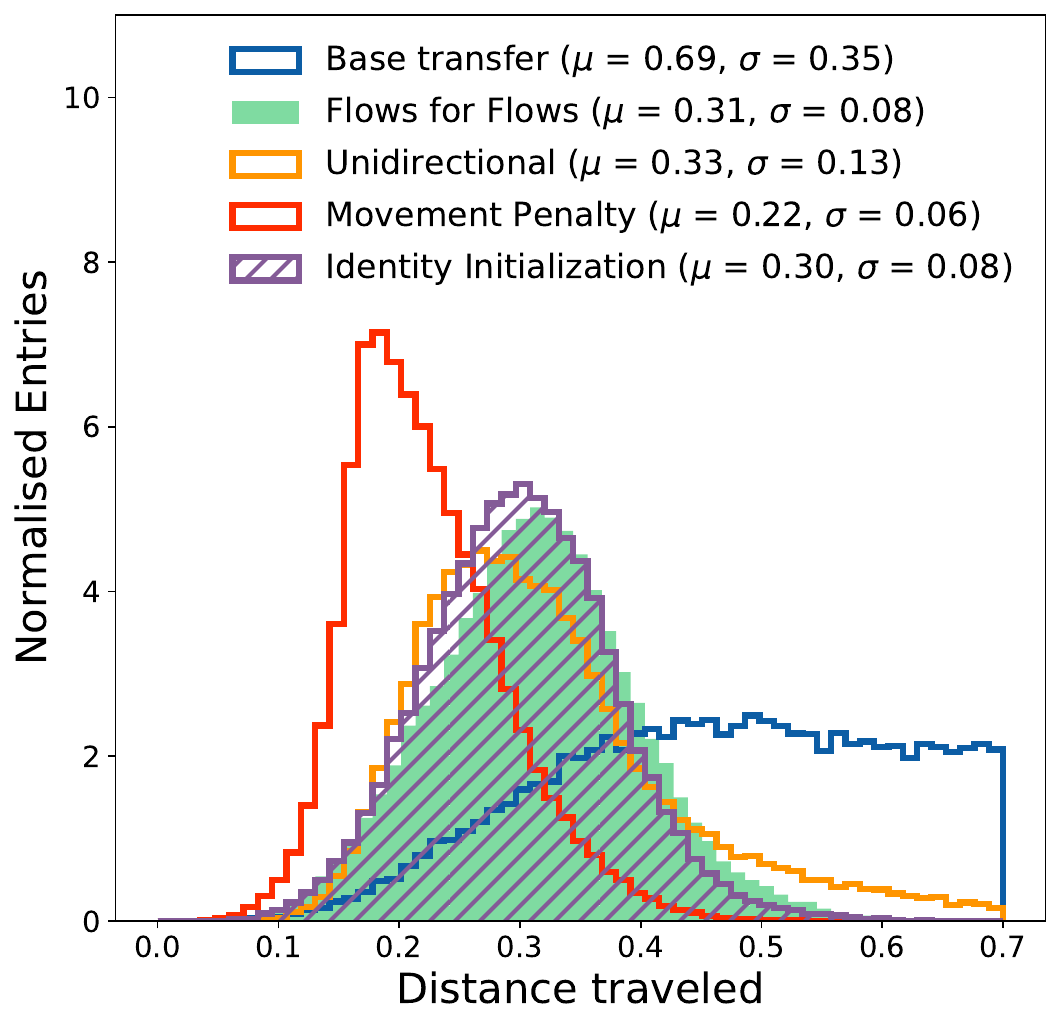}
    \caption{Distances traveled in 5-dimensional parameter space between the reference (less-than ideal auxiliary) dataset to the target (ideal auxiliary) dataset, outside of the signal region. The maximum possible distance travelable in parameter space (for the minmaxscaled features) is 13.41.}
    \label{fig:science_distances_traveled_total}
\end{figure}

\begin{figure*}
    \centering
    \includegraphics[width=\textwidth]{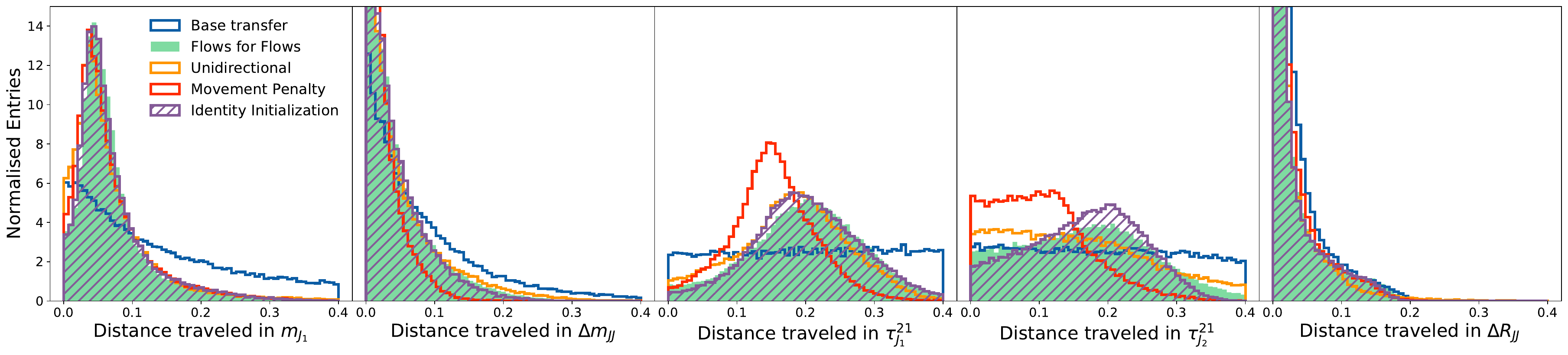}
    \caption{Distances traveled in each dimension of the 5-dimensional parameter space between the reference (less-than ideal auxiliary) dataset to the target (ideal auxiliary) dataset, outside of the signal region. The maximum possible distance travelable in each dimension of parameter space (for the minmaxscaled features) is 6.}
    \label{fig:science_distances_traveled_features}
\end{figure*}

In Fig.~\ref{fig:science_distances_traveled_total}, we show a histogram of the distances traveled for each data point due to the flow action. Distributions for distance traveled in each individual dimension of feature space are given in Fig.~\ref{fig:science_distances_traveled_features}. Since the reference and target distributions are so similar, the base transfer methods leads to a highly non-minimal transport path. While the unidirectional method performs well, it shows a longer tail in distance traveled that may represent a less-than-ideal mapping. The flows for flows and identity initialization methods perform comparably with relatively little distance traveled, while movement penalty appears to have found a nearly minimal path.

Based on the closeness of the distributions of the reference and target in Fig.~\ref{fig:features_f4f}, we might hope for a mapping that morphs features $m_{J_1}$, $\Delta m_{JJ}$, and $\Delta R_{JJ}$ almost not at all, and features $\tau^{21}_{J_1}$ and $\tau^{21}_{J_2}$ very minimally. Indeed, this is exactly the behavior we see in Fig.~\ref{fig:science_distances_traveled_features} for the movement penalty method (and, to a lesser extent, for the flows and flows and identity initialization methods).

     \section{Conclusions and future work}
 \label{sec:conclusions}

In this work, we have explored a number of ways to use normalizing flows to create mappings between nontrivial reference and target datasets of the same dimensionality. Our aim is to consider methods that go above and beyond the ``naive'' base transfer method, which uses standard normalizing flows that map from reference to target via a base density intermediary. In particular, we have introduced the flows for flows method, which uses two normalizing flows to parameterise the probability densities of both the reference and the target and trains both with exact maximum likelihoods.

We have evaluated five transfer methods: base transfer, unidirectional transfer, flows for flows, movement penalty, and identity initialization. We have attempted to evaluate each method on two facets: the accuracy of the transport between reference and target, and the efficiency of the transport (i.e. how far away are points transported by the mapping). When the reference and target are fully unrelated (such as for the toy examples in Sec.~\ref{sec:results_toy}), the flows for flows method is comparable with the naive base transfer method both for accuracy and extent. When the reference and target sets are similar, or obviously related in some way (such as for the particle physics calibration application in Sec.~\ref{sec:science}), the flows and flows method is far preferable to the base transfer method. These results imply that the flows for flows method should be used over the base transfer method, as it can always provide both an accurate and efficient transport. However, the highest performing (and thus our recommended) methods of transport are either the movement penalty or identity initialization methods, depending on the specific application.

There are many avenues for further modifications of the flows for flows method, or other ways to construct flow-based mapping functions in general. One interesting avenue involves physically-motivated extensions of normalizing flows: continuous normalizing flows (CNF) \cite{https://doi.org/10.48550/arxiv.1806.07366}, we can constrain the flow mappings such that they can be assigned velocity vectors, and convex potential (CP) flows \cite{cp_flows}, where the map is constrained to be the gradient of a convex potential. One can explicitly enforce optimal transports with OT-Flows \cite{DBLP:journals/corr/abs-2006-00104}, which add to the CNF loss both an L2 movement penalty and a penalty that encourages the mapping to transport points along the minimum of some potential function. While such modifications may not be necessary when the reference and target distributions are very similar, they could be explored for situations when the reference and target distributions are significantly different.




    \section*{Code}
    The flows for flows package can be found at \url{https://github.com/jraine/flows4flows}.  JR and SK contributed equally to its creation.

    \section*{Acknowledgements}
    TG, SK, and JR would like to acknowledge funding through the SNSF Sinergia grant called Robust Deep Density Models for High-Energy Particle Physics and Solar Flare Analysis (RODEM) with funding number CRSII$5\_193716$. BN and RM are supported by the U.S. Department of Energy (DOE), Office of Science under contract DE-AC02-05CH11231. RM is additionally supported by the National Science Foundation Graduate Research Fellowship Program under Grant No. DGE 2146752; any opinions, findings, and conclusions or recommendations expressed in this material are those of the authors and do not necessarily reflect the views of the National Science Foundation.
    \phantomsection
    \addcontentsline{toc}{chapter}{References}
    
    \bibliography{main}

\begin{thebibliography}{29}%
\makeatletter
\providecommand \@ifxundefined [1]{%
 \@ifx{#1\undefined}
}%
\providecommand \@ifnum [1]{%
 \ifnum #1\expandafter \@firstoftwo
 \else \expandafter \@secondoftwo
 \fi
}%
\providecommand \@ifx [1]{%
 \ifx #1\expandafter \@firstoftwo
 \else \expandafter \@secondoftwo
 \fi
}%
\providecommand \natexlab [1]{#1}%
\providecommand \enquote  [1]{``#1''}%
\providecommand \bibnamefont  [1]{#1}%
\providecommand \bibfnamefont [1]{#1}%
\providecommand \citenamefont [1]{#1}%
\providecommand \href@noop [0]{\@secondoftwo}%
\providecommand \href [0]{\begingroup \@sanitize@url \@href}%
\providecommand \@href[1]{\@@startlink{#1}\@@href}%
\providecommand \@@href[1]{\endgroup#1\@@endlink}%
\providecommand \@sanitize@url [0]{\catcode `\\12\catcode `\$12\catcode
  `\&12\catcode `\#12\catcode `\^12\catcode `\_12\catcode `\%12\relax}%
\providecommand \@@startlink[1]{}%
\providecommand \@@endlink[0]{}%
\providecommand \url  [0]{\begingroup\@sanitize@url \@url }%
\providecommand \@url [1]{\endgroup\@href {#1}{\urlprefix }}%
\providecommand \urlprefix  [0]{URL }%
\providecommand \Eprint [0]{\href }%
\providecommand \doibase [0]{https://doi.org/}%
\providecommand \selectlanguage [0]{\@gobble}%
\providecommand \bibinfo  [0]{\@secondoftwo}%
\providecommand \bibfield  [0]{\@secondoftwo}%
\providecommand \translation [1]{[#1]}%
\providecommand \BibitemOpen [0]{}%
\providecommand \bibitemStop [0]{}%
\providecommand \bibitemNoStop [0]{.\EOS\space}%
\providecommand \EOS [0]{\spacefactor3000\relax}%
\providecommand \BibitemShut  [1]{\csname bibitem#1\endcsname}%
\let\auto@bib@innerbib\@empty
\bibitem [{\citenamefont {Hastie}\ \emph {et~al.}(2001)\citenamefont {Hastie},
  \citenamefont {Tibshirani},\ and\ \citenamefont
  {Friedman}}]{hastie01statisticallearning}%
  \BibitemOpen
  \bibfield  {author} {\bibinfo {author} {\bibfnamefont {T.}~\bibnamefont
  {Hastie}}, \bibinfo {author} {\bibfnamefont {R.}~\bibnamefont {Tibshirani}},\
  and\ \bibinfo {author} {\bibfnamefont {J.}~\bibnamefont {Friedman}},\
  }\href@noop {} {\emph {\bibinfo {title} {The Elements of Statistical
  Learning}}},\ Springer Series in Statistics\ (\bibinfo  {publisher} {Springer
  New York Inc.},\ \bibinfo {address} {New York, NY, USA},\ \bibinfo {year}
  {2001})\BibitemShut {NoStop}%
\bibitem [{\citenamefont {Sugiyama}\ \emph {et~al.}(2012)\citenamefont
  {Sugiyama}, \citenamefont {Suzuki},\ and\ \citenamefont
  {Kanamori}}]{sugiyama_suzuki_kanamori_2012}%
  \BibitemOpen
  \bibfield  {author} {\bibinfo {author} {\bibfnamefont {M.}~\bibnamefont
  {Sugiyama}}, \bibinfo {author} {\bibfnamefont {T.}~\bibnamefont {Suzuki}},\
  and\ \bibinfo {author} {\bibfnamefont {T.}~\bibnamefont {Kanamori}},\ }\href
  {https://doi.org/10.1017/CBO9781139035613} {\emph {\bibinfo {title} {Density
  Ratio Estimation in Machine Learning}}}\ (\bibinfo  {publisher} {Cambridge
  University Press},\ \bibinfo {year} {2012})\BibitemShut {NoStop}%
\bibitem [{\citenamefont {Tabak}\ and\ \citenamefont
  {Turner}(2013)}]{tabak_flows}%
  \BibitemOpen
  \bibfield  {author} {\bibinfo {author} {\bibfnamefont {E.~G.}\ \bibnamefont
  {Tabak}}\ and\ \bibinfo {author} {\bibfnamefont {C.~V.}\ \bibnamefont
  {Turner}},\ }\bibfield  {title} {\bibinfo {title} {A family of nonparametric
  density estimation algorithms},\ }\href
  {https://doi.org/https://doi.org/10.1002/cpa.21423} {\bibfield  {journal}
  {\bibinfo  {journal} {Communications on Pure and Applied Mathematics}\
  }\textbf {\bibinfo {volume} {66}},\ \bibinfo {pages} {145} (\bibinfo {year}
  {2013})},\ \Eprint
  {https://arxiv.org/abs/https://onlinelibrary.wiley.com/doi/pdf/10.1002/cpa.21423}
  {https://onlinelibrary.wiley.com/doi/pdf/10.1002/cpa.21423} \BibitemShut
  {NoStop}%
\bibitem [{\citenamefont {Papamakarios}\ \emph {et~al.}(2019)\citenamefont
  {Papamakarios}, \citenamefont {Nalisnick}, \citenamefont {Rezende},
  \citenamefont {Mohamed},\ and\ \citenamefont
  {Lakshminarayanan}}]{flows_review}%
  \BibitemOpen
  \bibfield  {author} {\bibinfo {author} {\bibfnamefont {G.}~\bibnamefont
  {Papamakarios}}, \bibinfo {author} {\bibfnamefont {E.}~\bibnamefont
  {Nalisnick}}, \bibinfo {author} {\bibfnamefont {D.~J.}\ \bibnamefont
  {Rezende}}, \bibinfo {author} {\bibfnamefont {S.}~\bibnamefont {Mohamed}},\
  and\ \bibinfo {author} {\bibfnamefont {B.}~\bibnamefont {Lakshminarayanan}},\
  }\bibfield  {title} {\bibinfo {title} {Normalizing flows for probabilistic
  modeling and inference}\ }\href {https://doi.org/10.48550/ARXIV.1912.02762}
  {10.48550/ARXIV.1912.02762} (\bibinfo {year} {2019})\BibitemShut {NoStop}%
\bibitem [{\citenamefont {Raine}\ \emph {et~al.}(2023)\citenamefont {Raine},
  \citenamefont {Klein}, \citenamefont {Sengupta},\ and\ \citenamefont
  {Golling}}]{Raine:2022hht}%
  \BibitemOpen
  \bibfield  {author} {\bibinfo {author} {\bibfnamefont {J.~A.}\ \bibnamefont
  {Raine}}, \bibinfo {author} {\bibfnamefont {S.}~\bibnamefont {Klein}},
  \bibinfo {author} {\bibfnamefont {D.}~\bibnamefont {Sengupta}},\ and\
  \bibinfo {author} {\bibfnamefont {T.}~\bibnamefont {Golling}},\ }\bibfield
  {title} {\bibinfo {title} {{CURTAINs for your sliding window: Constructing
  unobserved regions by transforming adjacent intervals}},\ }\href
  {https://doi.org/10.3389/fdata.2023.899345} {\bibfield  {journal} {\bibinfo
  {journal} {Front. Big Data}\ }\textbf {\bibinfo {volume} {6}},\ \bibinfo
  {pages} {899345} (\bibinfo {year} {2023})},\ \Eprint
  {https://arxiv.org/abs/2203.09470} {arXiv:2203.09470 [hep-ph]} \BibitemShut
  {NoStop}%
\bibitem [{\citenamefont {Golling}\ \emph {et~al.}(2023)\citenamefont
  {Golling}, \citenamefont {Klein}, \citenamefont {Mastandrea},\ and\
  \citenamefont {Nachman}}]{Golling:2022nkl}%
  \BibitemOpen
  \bibfield  {author} {\bibinfo {author} {\bibfnamefont {T.}~\bibnamefont
  {Golling}}, \bibinfo {author} {\bibfnamefont {S.}~\bibnamefont {Klein}},
  \bibinfo {author} {\bibfnamefont {R.}~\bibnamefont {Mastandrea}},\ and\
  \bibinfo {author} {\bibfnamefont {B.}~\bibnamefont {Nachman}},\ }\bibfield
  {title} {\bibinfo {title} {{Flow-enhanced transportation for anomaly
  detection}},\ }\href {https://doi.org/10.1103/PhysRevD.107.096025} {\bibfield
   {journal} {\bibinfo  {journal} {Phys. Rev. D}\ }\textbf {\bibinfo {volume}
  {107}},\ \bibinfo {pages} {096025} (\bibinfo {year} {2023})},\ \Eprint
  {https://arxiv.org/abs/2212.11285} {arXiv:2212.11285 [hep-ph]} \BibitemShut
  {NoStop}%
\bibitem [{\citenamefont {Sengupta}\ \emph {et~al.}(2023)\citenamefont
  {Sengupta}, \citenamefont {Klein}, \citenamefont {Raine},\ and\ \citenamefont
  {Golling}}]{Sengupta:2023xqy}%
  \BibitemOpen
  \bibfield  {author} {\bibinfo {author} {\bibfnamefont {D.}~\bibnamefont
  {Sengupta}}, \bibinfo {author} {\bibfnamefont {S.}~\bibnamefont {Klein}},
  \bibinfo {author} {\bibfnamefont {J.~A.}\ \bibnamefont {Raine}},\ and\
  \bibinfo {author} {\bibfnamefont {T.}~\bibnamefont {Golling}},\ }\bibfield
  {title} {\bibinfo {title} {{CURTAINs Flows For Flows: Constructing Unobserved
  Regions with Maximum Likelihood Estimation}},\ }\href@noop {} {\  (\bibinfo
  {year} {2023})},\ \Eprint {https://arxiv.org/abs/2305.04646}
  {arXiv:2305.04646 [hep-ph]} \BibitemShut {NoStop}%
\bibitem [{\citenamefont {Howard}\ \emph {et~al.}(2022)\citenamefont {Howard},
  \citenamefont {Mandt}, \citenamefont {Whiteson},\ and\ \citenamefont
  {Yang}}]{Howard:2021pos}%
  \BibitemOpen
  \bibfield  {author} {\bibinfo {author} {\bibfnamefont {J.~N.}\ \bibnamefont
  {Howard}}, \bibinfo {author} {\bibfnamefont {S.}~\bibnamefont {Mandt}},
  \bibinfo {author} {\bibfnamefont {D.}~\bibnamefont {Whiteson}},\ and\
  \bibinfo {author} {\bibfnamefont {Y.}~\bibnamefont {Yang}},\ }\bibfield
  {title} {\bibinfo {title} {{Learning to simulate high energy particle
  collisions from unlabeled data}},\ }\href
  {https://doi.org/10.1038/s41598-022-10966-7} {\bibfield  {journal} {\bibinfo
  {journal} {Sci. Rep.}\ }\textbf {\bibinfo {volume} {12}},\ \bibinfo {pages}
  {7567} (\bibinfo {year} {2022})},\ \Eprint {https://arxiv.org/abs/2101.08944}
  {arXiv:2101.08944 [hep-ph]} \BibitemShut {NoStop}%
\bibitem [{\citenamefont {Diefenbacher}\ \emph {et~al.}(2023)\citenamefont
  {Diefenbacher}, \citenamefont {Mikuni},\ and\ \citenamefont
  {Nachman}}]{Diefenbacher:2023flw}%
  \BibitemOpen
  \bibfield  {author} {\bibinfo {author} {\bibfnamefont {S.}~\bibnamefont
  {Diefenbacher}}, \bibinfo {author} {\bibfnamefont {V.}~\bibnamefont
  {Mikuni}},\ and\ \bibinfo {author} {\bibfnamefont {B.}~\bibnamefont
  {Nachman}},\ }\bibfield  {title} {\bibinfo {title} {{Refining Fast
  Calorimeter Simulations with a Schr\"odinger Bridge}},\ }\href@noop {} {\
  (\bibinfo {year} {2023})},\ \Eprint {https://arxiv.org/abs/2308.12339}
  {arXiv:2308.12339 [physics.ins-det]} \BibitemShut {NoStop}%
\bibitem [{\citenamefont {Nachman}\ and\ \citenamefont
  {Shih}(2020)}]{Nachman:2020lpy}%
  \BibitemOpen
  \bibfield  {author} {\bibinfo {author} {\bibfnamefont {B.}~\bibnamefont
  {Nachman}}\ and\ \bibinfo {author} {\bibfnamefont {D.}~\bibnamefont {Shih}},\
  }\bibfield  {title} {\bibinfo {title} {{Anomaly Detection with Density
  Estimation}},\ }\href {https://doi.org/10.1103/PhysRevD.101.075042}
  {\bibfield  {journal} {\bibinfo  {journal} {Phys. Rev. D}\ }\textbf {\bibinfo
  {volume} {101}},\ \bibinfo {pages} {075042} (\bibinfo {year} {2020})},\
  \Eprint {https://arxiv.org/abs/2001.04990} {arXiv:2001.04990 [hep-ph]}
  \BibitemShut {NoStop}%
\bibitem [{\citenamefont {Stein}\ \emph {et~al.}(2020)\citenamefont {Stein},
  \citenamefont {Seljak},\ and\ \citenamefont {Dai}}]{Stein:2020rou}%
  \BibitemOpen
  \bibfield  {author} {\bibinfo {author} {\bibfnamefont {G.}~\bibnamefont
  {Stein}}, \bibinfo {author} {\bibfnamefont {U.}~\bibnamefont {Seljak}},\ and\
  \bibinfo {author} {\bibfnamefont {B.}~\bibnamefont {Dai}},\ }\bibfield
  {title} {\bibinfo {title} {{Unsupervised in-distribution anomaly detection of
  new physics through conditional density estimation}},\ }in\ \href@noop {}
  {\emph {\bibinfo {booktitle} {{34th Conference on Neural Information
  Processing Systems}}}}\ (\bibinfo {year} {2020})\ \Eprint
  {https://arxiv.org/abs/2012.11638} {arXiv:2012.11638 [cs.LG]} \BibitemShut
  {NoStop}%
\bibitem [{\citenamefont {Hallin}\ \emph {et~al.}(2021)\citenamefont {Hallin},
  \citenamefont {Isaacson}, \citenamefont {Kasieczka}, \citenamefont {Krause},
  \citenamefont {Nachman}, \citenamefont {Quadfasel}, \citenamefont
  {Schlaffer}, \citenamefont {Shih},\ and\ \citenamefont
  {Sommerhalder}}]{hallin2021classifying}%
  \BibitemOpen
  \bibfield  {author} {\bibinfo {author} {\bibfnamefont {A.}~\bibnamefont
  {Hallin}}, \bibinfo {author} {\bibfnamefont {J.}~\bibnamefont {Isaacson}},
  \bibinfo {author} {\bibfnamefont {G.}~\bibnamefont {Kasieczka}}, \bibinfo
  {author} {\bibfnamefont {C.}~\bibnamefont {Krause}}, \bibinfo {author}
  {\bibfnamefont {B.}~\bibnamefont {Nachman}}, \bibinfo {author} {\bibfnamefont
  {T.}~\bibnamefont {Quadfasel}}, \bibinfo {author} {\bibfnamefont
  {M.}~\bibnamefont {Schlaffer}}, \bibinfo {author} {\bibfnamefont
  {D.}~\bibnamefont {Shih}},\ and\ \bibinfo {author} {\bibfnamefont
  {M.}~\bibnamefont {Sommerhalder}},\ }\bibfield  {title} {\bibinfo {title}
  {Classifying anomalies through outer density estimation (cathode)},\
  }\href@noop {} {\bibfield  {journal} {\bibinfo  {journal} {arXiv preprint
  arXiv:2109.00546}\ } (\bibinfo {year} {2021})}\BibitemShut {NoStop}%
\bibitem [{\citenamefont {Durkan}\ \emph {et~al.}(2019)\citenamefont {Durkan},
  \citenamefont {Bekasov}, \citenamefont {Murray},\ and\ \citenamefont
  {Papamakarios}}]{durkan2019neural}%
  \BibitemOpen
  \bibfield  {author} {\bibinfo {author} {\bibfnamefont {C.}~\bibnamefont
  {Durkan}}, \bibinfo {author} {\bibfnamefont {A.}~\bibnamefont {Bekasov}},
  \bibinfo {author} {\bibfnamefont {I.}~\bibnamefont {Murray}},\ and\ \bibinfo
  {author} {\bibfnamefont {G.}~\bibnamefont {Papamakarios}},\ }\bibfield
  {title} {\bibinfo {title} {Neural spline flows},\ }\href@noop {} {\bibfield
  {journal} {\bibinfo  {journal} {Advances in neural information processing
  systems}\ }\textbf {\bibinfo {volume} {32}} (\bibinfo {year}
  {2019})}\BibitemShut {NoStop}%
\bibitem [{\citenamefont {Durkan}\ \emph {et~al.}(2020)\citenamefont {Durkan},
  \citenamefont {Bekasov}, \citenamefont {Murray},\ and\ \citenamefont
  {Papamakarios}}]{nflows}%
  \BibitemOpen
  \bibfield  {author} {\bibinfo {author} {\bibfnamefont {C.}~\bibnamefont
  {Durkan}}, \bibinfo {author} {\bibfnamefont {A.}~\bibnamefont {Bekasov}},
  \bibinfo {author} {\bibfnamefont {I.}~\bibnamefont {Murray}},\ and\ \bibinfo
  {author} {\bibfnamefont {G.}~\bibnamefont {Papamakarios}},\ }\href
  {https://doi.org/10.5281/zenodo.4296287} {\bibinfo {title} {{nflows}:
  normalizing flows in {PyTorch}}} (\bibinfo {year} {2020})\BibitemShut
  {NoStop}%
\bibitem [{\citenamefont {Loshchilov}\ and\ \citenamefont
  {Hutter}(2016)}]{cosine_annealing}%
  \BibitemOpen
  \bibfield  {author} {\bibinfo {author} {\bibfnamefont {I.}~\bibnamefont
  {Loshchilov}}\ and\ \bibinfo {author} {\bibfnamefont {F.}~\bibnamefont
  {Hutter}},\ }\bibfield  {title} {\bibinfo {title} {Sgdr: Stochastic gradient
  descent with warm restarts},\ }\href@noop {} {\bibfield  {journal} {\bibinfo
  {journal} {arXiv preprint arXiv:1608.03983}\ } (\bibinfo {year}
  {2016})}\BibitemShut {NoStop}%
\bibitem [{\citenamefont {Karagiorgi}\ \emph {et~al.}(2021)\citenamefont
  {Karagiorgi}, \citenamefont {Kasieczka}, \citenamefont {Kravitz},
  \citenamefont {Nachman},\ and\ \citenamefont {Shih}}]{Karagiorgi:2021ngt}%
  \BibitemOpen
  \bibfield  {author} {\bibinfo {author} {\bibfnamefont {G.}~\bibnamefont
  {Karagiorgi}}, \bibinfo {author} {\bibfnamefont {G.}~\bibnamefont
  {Kasieczka}}, \bibinfo {author} {\bibfnamefont {S.}~\bibnamefont {Kravitz}},
  \bibinfo {author} {\bibfnamefont {B.}~\bibnamefont {Nachman}},\ and\ \bibinfo
  {author} {\bibfnamefont {D.}~\bibnamefont {Shih}},\ }\bibfield  {title}
  {\bibinfo {title} {{Machine Learning in the Search for New Fundamental
  Physics}},\ }\href {https://doi.org/10.1038/s42254-022-00455-1} {\bibfield
  {journal} {\bibinfo  {journal} {Nature Reviews Physics}\ }\textbf {\bibinfo
  {volume} {4}},\ \bibinfo {pages} {399} (\bibinfo {year} {2021})},\ \Eprint
  {https://arxiv.org/abs/2112.03769} {arXiv:2112.03769 [hep-ph]} \BibitemShut
  {NoStop}%
\bibitem [{\citenamefont {Kasieczka}\ \emph
  {et~al.}(2021{\natexlab{a}})\citenamefont {Kasieczka}, \citenamefont
  {Nachman},\ and\ \citenamefont {Shih}}]{Kasieczka:2021tew}%
  \BibitemOpen
  \bibfield  {author} {\bibinfo {author} {\bibfnamefont {G.}~\bibnamefont
  {Kasieczka}}, \bibinfo {author} {\bibfnamefont {B.}~\bibnamefont {Nachman}},\
  and\ \bibinfo {author} {\bibfnamefont {D.}~\bibnamefont {Shih}},\ }\bibfield
  {title} {\bibinfo {title} {{New Methods and Datasets for Group Anomaly
  Detection From Fundamental Physics}},\ }in\ \href@noop {} {\emph {\bibinfo
  {booktitle} {{Conference on Knowledge Discovery and Data Mining}}}}\
  (\bibinfo {year} {2021})\ \Eprint {https://arxiv.org/abs/2107.02821}
  {arXiv:2107.02821 [stat.ML]} \BibitemShut {NoStop}%
\bibitem [{\citenamefont {Kasieczka}\ \emph {et~al.}(2019)\citenamefont
  {Kasieczka}, \citenamefont {Nachman},\ and\ \citenamefont
  {Shih}}]{LHCOlympics}%
  \BibitemOpen
  \bibfield  {author} {\bibinfo {author} {\bibfnamefont {G.}~\bibnamefont
  {Kasieczka}}, \bibinfo {author} {\bibfnamefont {B.}~\bibnamefont {Nachman}},\
  and\ \bibinfo {author} {\bibfnamefont {D.}~\bibnamefont {Shih}},\ }\href@noop
  {} {\bibinfo {title} {{Official Datasets for LHC Olympics 2020 Anomaly
  Detection Challenge (Version v6) [Data set].}}} (\bibinfo {year} {2019}),\
  \bibinfo {note} {\url{https://doi.org/10.5281/zenodo.4536624}}\BibitemShut
  {NoStop}%
\bibitem [{\citenamefont {Kasieczka}\ \emph
  {et~al.}(2021{\natexlab{b}})\citenamefont {Kasieczka} \emph
  {et~al.}}]{Kasieczka:2021xcg}%
  \BibitemOpen
  \bibfield  {author} {\bibinfo {author} {\bibfnamefont {G.}~\bibnamefont
  {Kasieczka}} \emph {et~al.},\ }\bibfield  {title} {\bibinfo {title} {{The LHC
  Olympics 2020 a community challenge for anomaly detection in high energy
  physics}},\ }\href {https://doi.org/10.1088/1361-6633/ac36b9} {\bibfield
  {journal} {\bibinfo  {journal} {Rept. Prog. Phys.}\ }\textbf {\bibinfo
  {volume} {84}},\ \bibinfo {pages} {124201} (\bibinfo {year}
  {2021}{\natexlab{b}})},\ \Eprint {https://arxiv.org/abs/2101.08320}
  {arXiv:2101.08320 [hep-ph]} \BibitemShut {NoStop}%
\bibitem [{\citenamefont {Cacciari}\ \emph {et~al.}(2012)\citenamefont
  {Cacciari}, \citenamefont {Salam},\ and\ \citenamefont
  {Soyez}}]{Cacciari:2011ma}%
  \BibitemOpen
  \bibfield  {author} {\bibinfo {author} {\bibfnamefont {M.}~\bibnamefont
  {Cacciari}}, \bibinfo {author} {\bibfnamefont {G.~P.}\ \bibnamefont
  {Salam}},\ and\ \bibinfo {author} {\bibfnamefont {G.}~\bibnamefont {Soyez}},\
  }\bibfield  {title} {\bibinfo {title} {{FastJet User Manual}},\ }\href
  {https://doi.org/10.1140/epjc/s10052-012-1896-2} {\bibfield  {journal}
  {\bibinfo  {journal} {Eur. Phys. J. C}\ }\textbf {\bibinfo {volume} {72}},\
  \bibinfo {pages} {1896} (\bibinfo {year} {2012})},\ \Eprint
  {https://arxiv.org/abs/1111.6097} {arXiv:1111.6097 [hep-ph]} \BibitemShut
  {NoStop}%
\bibitem [{\citenamefont {Cacciari}\ and\ \citenamefont
  {Salam}(2006)}]{Cacciari:2005hq}%
  \BibitemOpen
  \bibfield  {author} {\bibinfo {author} {\bibfnamefont {M.}~\bibnamefont
  {Cacciari}}\ and\ \bibinfo {author} {\bibfnamefont {G.~P.}\ \bibnamefont
  {Salam}},\ }\bibfield  {title} {\bibinfo {title} {{Dispelling the $N^{3}$
  myth for the $k_t$ jet-finder}},\ }\href
  {https://doi.org/10.1016/j.physletb.2006.08.037} {\bibfield  {journal}
  {\bibinfo  {journal} {Phys. Lett. B}\ }\textbf {\bibinfo {volume} {641}},\
  \bibinfo {pages} {57} (\bibinfo {year} {2006})},\ \Eprint
  {https://arxiv.org/abs/hep-ph/0512210} {arXiv:hep-ph/0512210} \BibitemShut
  {NoStop}%
\bibitem [{\citenamefont {Cacciari}\ \emph {et~al.}(2008)\citenamefont
  {Cacciari}, \citenamefont {Salam},\ and\ \citenamefont
  {Soyez}}]{Cacciari:2008gp}%
  \BibitemOpen
  \bibfield  {author} {\bibinfo {author} {\bibfnamefont {M.}~\bibnamefont
  {Cacciari}}, \bibinfo {author} {\bibfnamefont {G.~P.}\ \bibnamefont
  {Salam}},\ and\ \bibinfo {author} {\bibfnamefont {G.}~\bibnamefont {Soyez}},\
  }\bibfield  {title} {\bibinfo {title} {{The anti-$k_t$ jet clustering
  algorithm}},\ }\href {https://doi.org/10.1088/1126-6708/2008/04/063}
  {\bibfield  {journal} {\bibinfo  {journal} {JHEP}\ }\textbf {\bibinfo
  {volume} {04}},\ \bibinfo {pages} {063}},\ \Eprint
  {https://arxiv.org/abs/0802.1189} {arXiv:0802.1189 [hep-ph]} \BibitemShut
  {NoStop}%
\bibitem [{\citenamefont {Sjostrand}\ \emph {et~al.}(2006)\citenamefont
  {Sjostrand}, \citenamefont {Mrenna},\ and\ \citenamefont
  {Skands}}]{Sjostrand:2006za}%
  \BibitemOpen
  \bibfield  {author} {\bibinfo {author} {\bibfnamefont {T.}~\bibnamefont
  {Sjostrand}}, \bibinfo {author} {\bibfnamefont {S.}~\bibnamefont {Mrenna}},\
  and\ \bibinfo {author} {\bibfnamefont {P.~Z.}\ \bibnamefont {Skands}},\
  }\bibfield  {title} {\bibinfo {title} {{PYTHIA 6.4 Physics and Manual}},\
  }\href {https://doi.org/10.1088/1126-6708/2006/05/026} {\bibfield  {journal}
  {\bibinfo  {journal} {JHEP}\ }\textbf {\bibinfo {volume} {05}},\ \bibinfo
  {pages} {026}},\ \Eprint {https://arxiv.org/abs/hep-ph/0603175}
  {arXiv:hep-ph/0603175} \BibitemShut {NoStop}%
\bibitem [{\citenamefont {Sj\"ostrand}\ \emph {et~al.}(2015)\citenamefont
  {Sj\"ostrand}, \citenamefont {Ask}, \citenamefont {Christiansen},
  \citenamefont {Corke}, \citenamefont {Desai}, \citenamefont {Ilten},
  \citenamefont {Mrenna}, \citenamefont {Prestel}, \citenamefont {Rasmussen},\
  and\ \citenamefont {Skands}}]{Sjostrand:2014zea}%
  \BibitemOpen
  \bibfield  {author} {\bibinfo {author} {\bibfnamefont {T.}~\bibnamefont
  {Sj\"ostrand}}, \bibinfo {author} {\bibfnamefont {S.}~\bibnamefont {Ask}},
  \bibinfo {author} {\bibfnamefont {J.~R.}\ \bibnamefont {Christiansen}},
  \bibinfo {author} {\bibfnamefont {R.}~\bibnamefont {Corke}}, \bibinfo
  {author} {\bibfnamefont {N.}~\bibnamefont {Desai}}, \bibinfo {author}
  {\bibfnamefont {P.}~\bibnamefont {Ilten}}, \bibinfo {author} {\bibfnamefont
  {S.}~\bibnamefont {Mrenna}}, \bibinfo {author} {\bibfnamefont
  {S.}~\bibnamefont {Prestel}}, \bibinfo {author} {\bibfnamefont {C.~O.}\
  \bibnamefont {Rasmussen}},\ and\ \bibinfo {author} {\bibfnamefont {P.~Z.}\
  \bibnamefont {Skands}},\ }\bibfield  {title} {\bibinfo {title} {{An
  introduction to PYTHIA 8.2}},\ }\href
  {https://doi.org/10.1016/j.cpc.2015.01.024} {\bibfield  {journal} {\bibinfo
  {journal} {Comput. Phys. Commun.}\ }\textbf {\bibinfo {volume} {191}},\
  \bibinfo {pages} {159} (\bibinfo {year} {2015})},\ \Eprint
  {https://arxiv.org/abs/1410.3012} {arXiv:1410.3012 [hep-ph]} \BibitemShut
  {NoStop}%
\bibitem [{\citenamefont {Bähr}\ \emph {et~al.}(2008)\citenamefont {Bähr},
  \citenamefont {Gieseke}, \citenamefont {Gigg}, \citenamefont {Grellscheid},
  \citenamefont {Hamilton}, \citenamefont {Latunde-Dada}, \citenamefont
  {Plätzer}, \citenamefont {Richardson}, \citenamefont {Seymour},
  \citenamefont {Sherstnev},\ and\ \citenamefont {Webber}}]{B_hr_2008}%
  \BibitemOpen
  \bibfield  {author} {\bibinfo {author} {\bibfnamefont {M.}~\bibnamefont
  {Bähr}}, \bibinfo {author} {\bibfnamefont {S.}~\bibnamefont {Gieseke}},
  \bibinfo {author} {\bibfnamefont {M.~A.}\ \bibnamefont {Gigg}}, \bibinfo
  {author} {\bibfnamefont {D.}~\bibnamefont {Grellscheid}}, \bibinfo {author}
  {\bibfnamefont {K.}~\bibnamefont {Hamilton}}, \bibinfo {author}
  {\bibfnamefont {O.}~\bibnamefont {Latunde-Dada}}, \bibinfo {author}
  {\bibfnamefont {S.}~\bibnamefont {Plätzer}}, \bibinfo {author}
  {\bibfnamefont {P.}~\bibnamefont {Richardson}}, \bibinfo {author}
  {\bibfnamefont {M.~H.}\ \bibnamefont {Seymour}}, \bibinfo {author}
  {\bibfnamefont {A.}~\bibnamefont {Sherstnev}},\ and\ \bibinfo {author}
  {\bibfnamefont {B.~R.}\ \bibnamefont {Webber}},\ }\bibfield  {title}
  {\bibinfo {title} {Herwig++ physics and manual},\ }\href
  {https://doi.org/10.1140/epjc/s10052-008-0798-9} {\bibfield  {journal}
  {\bibinfo  {journal} {The European Physical Journal C}\ }\textbf {\bibinfo
  {volume} {58}},\ \bibinfo {pages} {639} (\bibinfo {year} {2008})}\BibitemShut
  {NoStop}%
\bibitem [{\citenamefont {Thaler}\ and\ \citenamefont
  {Van~Tilburg}(2011)}]{Thaler:2010tr}%
  \BibitemOpen
  \bibfield  {author} {\bibinfo {author} {\bibfnamefont {J.}~\bibnamefont
  {Thaler}}\ and\ \bibinfo {author} {\bibfnamefont {K.}~\bibnamefont
  {Van~Tilburg}},\ }\bibfield  {title} {\bibinfo {title} {{Identifying Boosted
  Objects with N-subjettiness}},\ }\href
  {https://doi.org/10.1007/JHEP03(2011)015} {\bibfield  {journal} {\bibinfo
  {journal} {JHEP}\ }\textbf {\bibinfo {volume} {03}},\ \bibinfo {pages}
  {015}},\ \Eprint {https://arxiv.org/abs/1011.2268} {arXiv:1011.2268 [hep-ph]}
  \BibitemShut {NoStop}%
\bibitem [{\citenamefont {Chen}\ \emph {et~al.}(2018)\citenamefont {Chen},
  \citenamefont {Rubanova}, \citenamefont {Bettencourt},\ and\ \citenamefont
  {Duvenaud}}]{https://doi.org/10.48550/arxiv.1806.07366}%
  \BibitemOpen
  \bibfield  {author} {\bibinfo {author} {\bibfnamefont {R.~T.~Q.}\
  \bibnamefont {Chen}}, \bibinfo {author} {\bibfnamefont {Y.}~\bibnamefont
  {Rubanova}}, \bibinfo {author} {\bibfnamefont {J.}~\bibnamefont
  {Bettencourt}},\ and\ \bibinfo {author} {\bibfnamefont {D.}~\bibnamefont
  {Duvenaud}},\ }\href {https://doi.org/10.48550/ARXIV.1806.07366} {\bibinfo
  {title} {Neural ordinary differential equations}} (\bibinfo {year}
  {2018})\BibitemShut {NoStop}%
\bibitem [{\citenamefont {Huang}\ \emph {et~al.}(2020)\citenamefont {Huang},
  \citenamefont {Chen}, \citenamefont {Tsirigotis},\ and\ \citenamefont
  {Courville}}]{cp_flows}%
  \BibitemOpen
  \bibfield  {author} {\bibinfo {author} {\bibfnamefont {C.-W.}\ \bibnamefont
  {Huang}}, \bibinfo {author} {\bibfnamefont {R.~T.~Q.}\ \bibnamefont {Chen}},
  \bibinfo {author} {\bibfnamefont {C.}~\bibnamefont {Tsirigotis}},\ and\
  \bibinfo {author} {\bibfnamefont {A.}~\bibnamefont {Courville}},\ }\href
  {https://doi.org/10.48550/ARXIV.2012.05942} {\bibinfo {title} {Convex
  potential flows: Universal probability distributions with optimal transport
  and convex optimization}} (\bibinfo {year} {2020})\BibitemShut {NoStop}%
\bibitem [{\citenamefont {Onken}\ \emph {et~al.}(2020)\citenamefont {Onken},
  \citenamefont {Fung}, \citenamefont {Li},\ and\ \citenamefont
  {Ruthotto}}]{DBLP:journals/corr/abs-2006-00104}%
  \BibitemOpen
  \bibfield  {author} {\bibinfo {author} {\bibfnamefont {D.}~\bibnamefont
  {Onken}}, \bibinfo {author} {\bibfnamefont {S.~W.}\ \bibnamefont {Fung}},
  \bibinfo {author} {\bibfnamefont {X.}~\bibnamefont {Li}},\ and\ \bibinfo
  {author} {\bibfnamefont {L.}~\bibnamefont {Ruthotto}},\ }\bibfield  {title}
  {\bibinfo {title} {Ot-flow: Fast and accurate continuous normalizing flows
  via optimal transport},\ }\href {https://arxiv.org/abs/2006.00104} {\bibfield
   {journal} {\bibinfo  {journal} {CoRR}\ }\textbf {\bibinfo {volume}
  {abs/2006.00104}} (\bibinfo {year} {2020})},\ \Eprint
  {https://arxiv.org/abs/2006.00104} {2006.00104} \BibitemShut {NoStop}%
\end{thebibliography}%
    
    \clearpage
    \onecolumngrid
    \appendix
    \counterwithin{figure}{section}

\section{Toy results: learning the identity}
\label{app:plots}

In Sec. \ref{sec:results_toy}, we focused on mappings between a nonidentical reference and target distribution. However, it is possible to get a more meaningful interpretation of a given transport method by considering an identical reference and target -- in other words, when we ask the flow to learn the identity mapping. In this situation, it would be desirable for the flow to learn to map data points as little as possible.

\begin{figure*}[h]
      \centering
    \includegraphics[width = .9\textwidth]{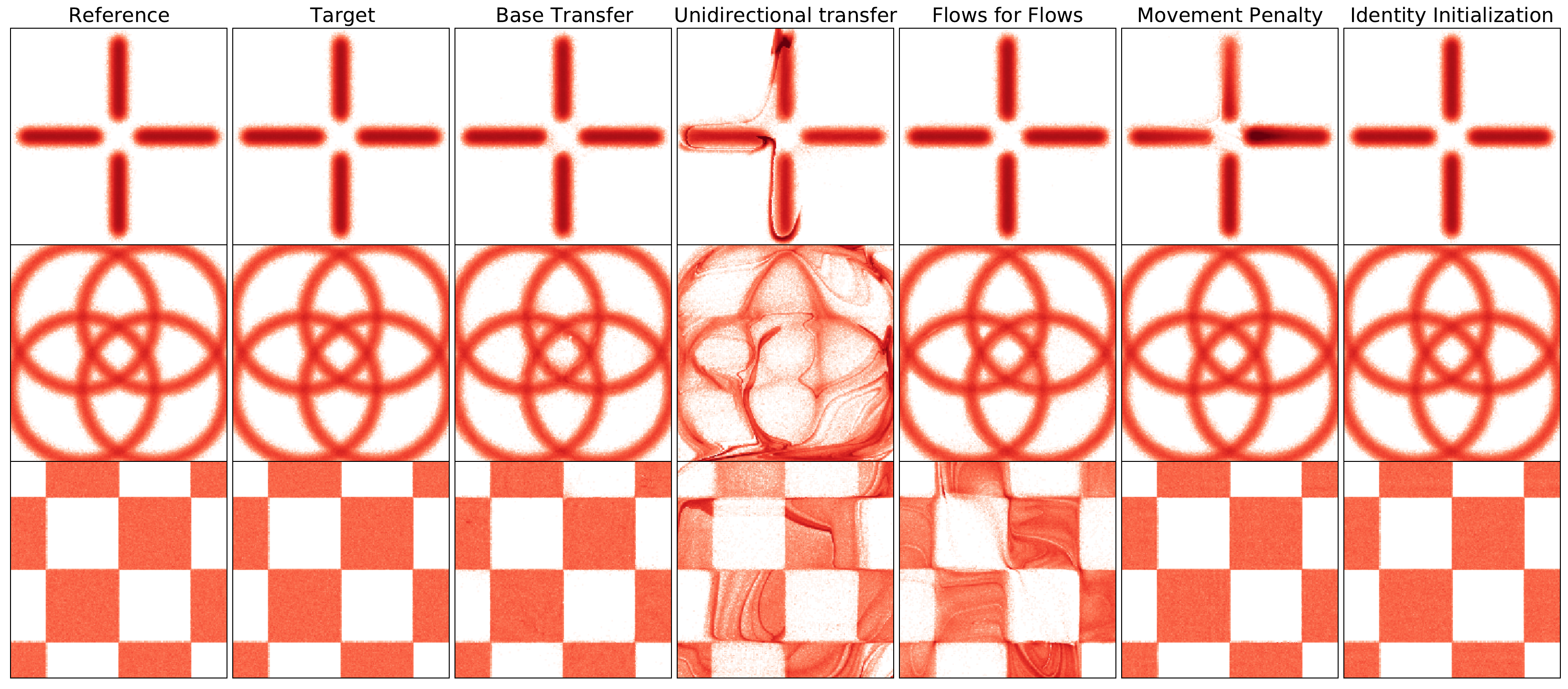}
    \caption{Transport tasks between various choices of identical reference and target toy distributions. The colorbar has been set to scale logarithmically, which can emphasize out-of-distribution points.}
    \label{fig:summary_identity}
\end{figure*}

In Fig.~\ref{fig:summary_identity}, we plot the transport results between three identical reference -- target parings (four overlapping circles, a four-pointed star, and a checkerboard pattern). All of the transfer methods shown are able to successfully learn to map from reference to target, except the unidirectional transfer, which fails glaringly. The base transfer, movement penalty, and identity initialization methods show the cleanest final-state distributions.

\begin{figure}[h]
    \centering
    \includegraphics[width=.365\textwidth]{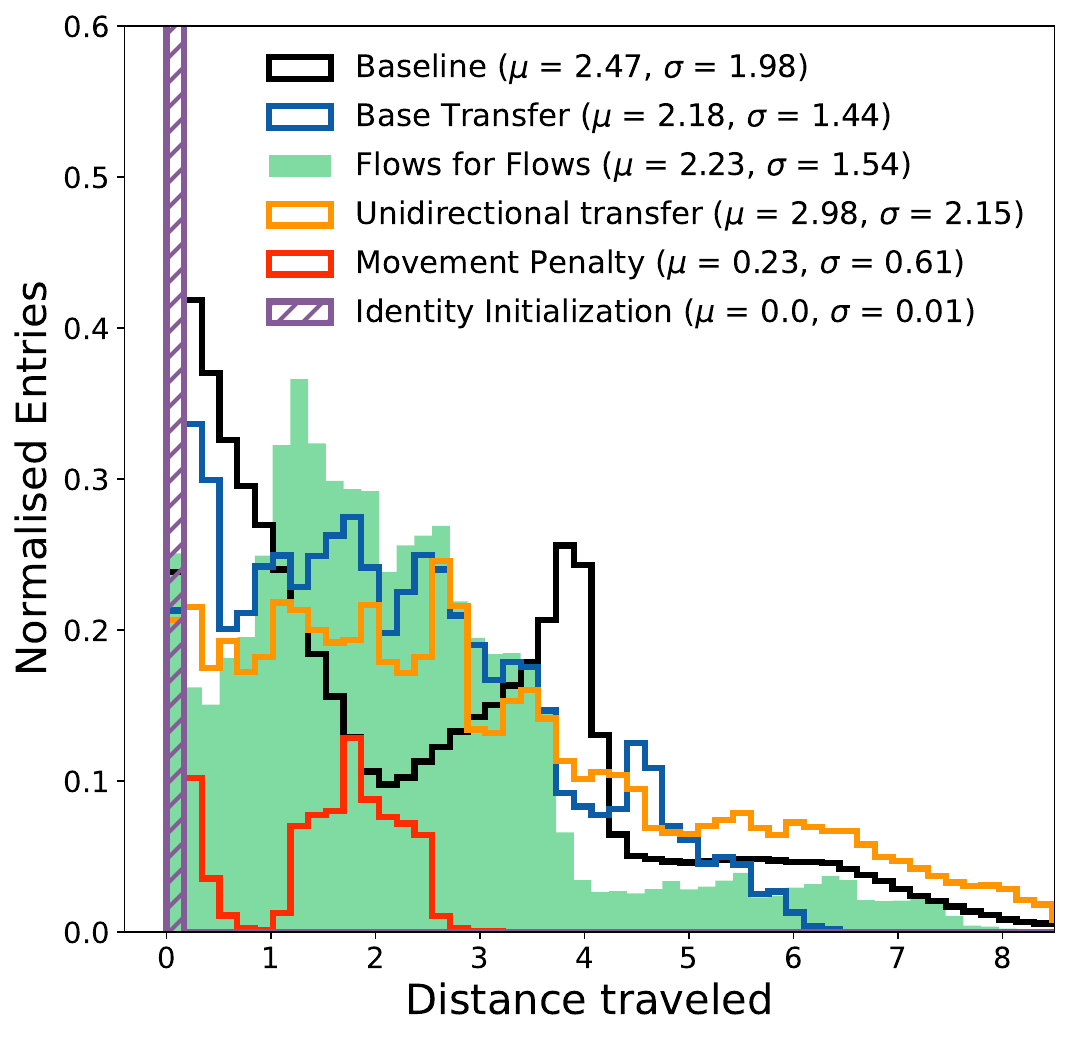}
    \caption{Distances traveled in parameter space between two identical toy distributions. Each histogram contains data from three transfer tasks, corresponding to the rows of Fig.~\ref{fig:summary_identity}.}
    \label{fig:distances_traveled_identity}
\end{figure}

In Fig.~\ref{fig:distances_traveled_identity}, we provide a compiled histogram of the distances traveled for the three transfer tasks shown in the rows of Fig.~\ref{fig:summary_identity}. In this case, the identity initialization method performs ideally, virtually leaving the reference distribution unchanged. The movement penalty method also performs well. The base transfer and flows for flows method are comparable and perform about as well as, if not slightly better than, the baseline (as was the case for nonidentical reference and target distributions); the unidirectional method performs suboptimally. 

    \section{Conditional toy distributions}
\label{app:conditional_dists}

In this appendix, we provide plots similar to those in Sec.~\ref{sec:results_toy} but for conditional normalizing flows (as given in Sec.~\ref{sec:nf_transfer}). We extend the two-dimensional toy distributions that have been studied so far by introducing rotations. A flows for flows model can then learn to move points sampled at one value of the condition such that they follow the distribution defined by another value of the condition. 

To train the flows for flows models for the conditional toy datasets: we first generate unconditional toy distributions. We then rotate each data point in the distribution $x_i$ by a random angle $\theta_i \in [7.5^\circ, 15^\circ, 22.5^\circ, 30^\circ, 37.5^\circ, 45^\circ]$. For a given reference-target data point pairing $x_R$, $x_T$ used to train the flow, the condition is then set to be the difference between the training data's rotation angles, $\theta_R - \theta_T$. Otherwise, the training procedure is the same as that of the unconditional toy distributions.

\begin{figure*}[h]
      \centering
    \includegraphics[width =.8\textwidth]{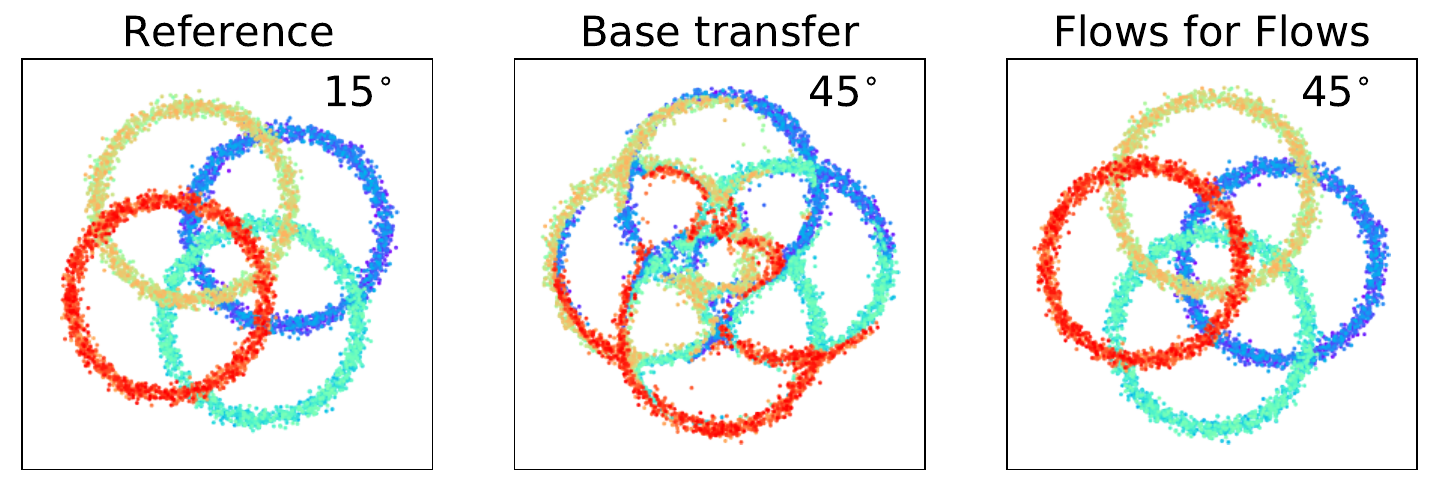}
    \caption{Transport task between a choice of identical reference and target toy distributions. Individual samples have been color coded so as to make clear their paths assigned by the transport method. The conditioning rotation of each distribution is given in the top right corner.}\label{fig:fourcircles_fourcircles_cond_base_transfer_f4f_15_00_45_00}
\end{figure*}

\begin{figure*}
      \centering
    \includegraphics[width = \textwidth]{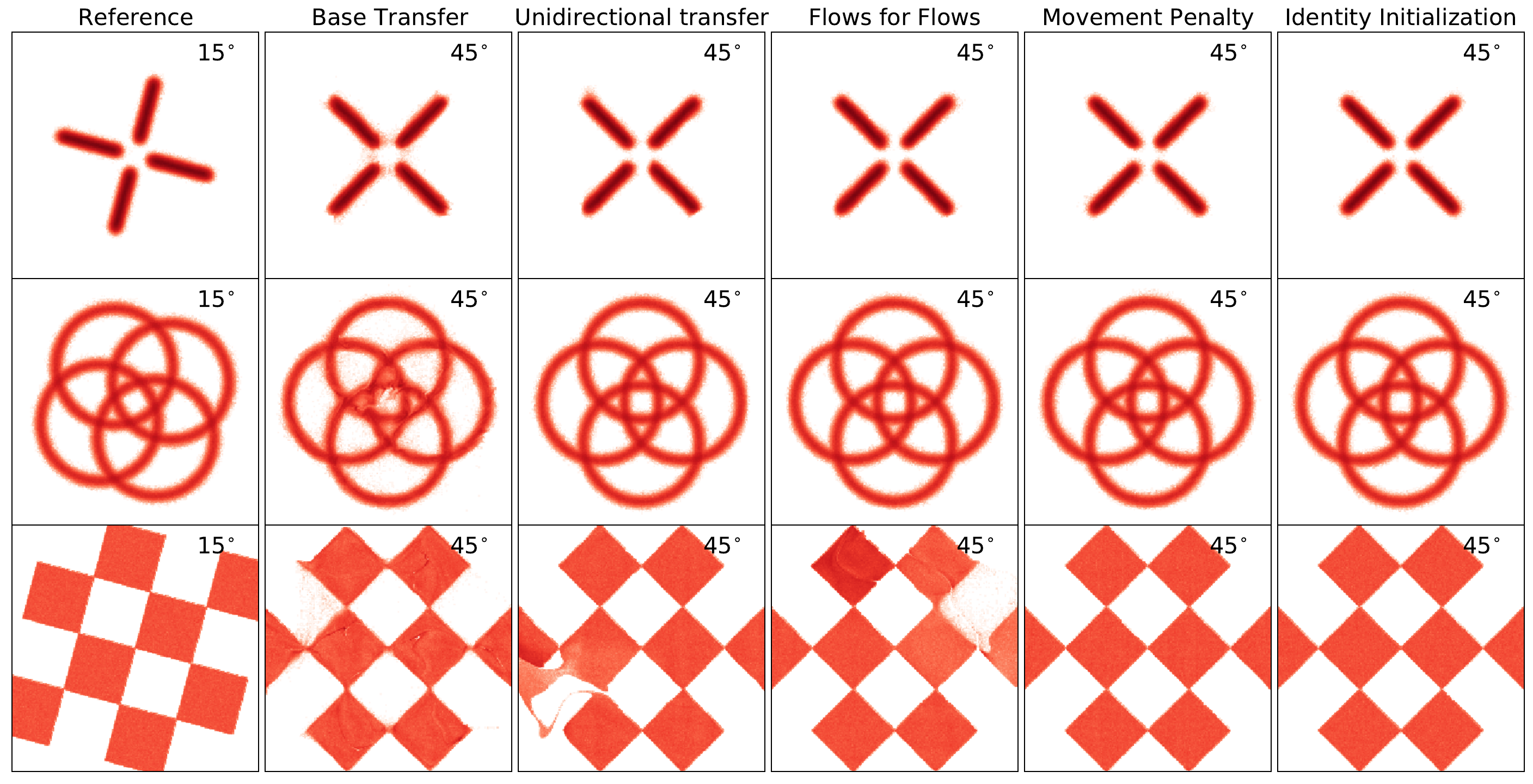}
    \caption{Transport tasks between various choices of identical reference and target toy distributions. The colorbar
has been set to scale logarithmically, which can emphasize out-of-distribution points.}
    \label{fig:summary_cond_identity_15_00_45_00}
\end{figure*}

The results for the conditional toy distribution transportation tasks tell a similar story as before, in Sec. \ref{sec:results_toy}. For a transport task between identical toy distributions with different conditioning angles (shown in Fig. \ref{fig:fourcircles_fourcircles_cond_base_transfer_f4f_15_00_45_00}), the flows for flows method appears to minimally and logically transport individual samples, in contrast to the base transport method. When looking more broadly at just the transport accuracy (shown in Fig. \ref{fig:summary_cond_identity_15_00_45_00}), the flows for flows-based methods (i.e. flows for flows, movement penalty, and identity initialization) are visibly better than the base transfer and unidirectional transfer.

\begin{figure*}[h]
      \centering
    \includegraphics[width = .8\textwidth]{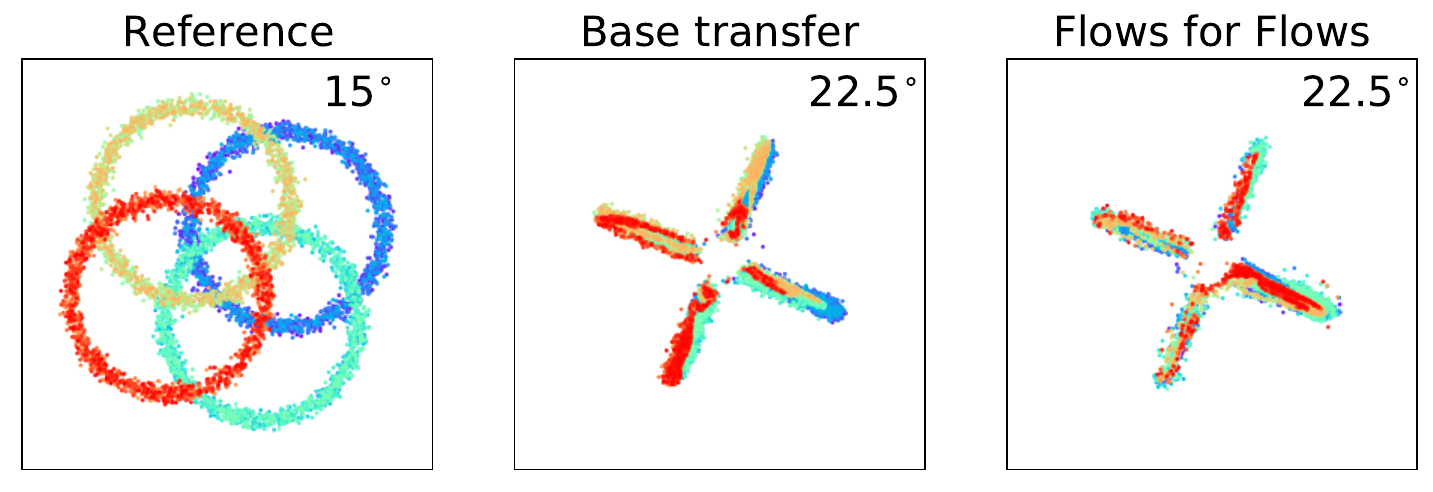}
    \caption{Transport task between a nonidentical choice of reference and target toy distributions. Individual samples have been color coded so as to make clear their paths assigned by the transport method.}
\label{fig:fourcircles_star_cond_base_transfer_f4f_15_00_22_50}
\end{figure*}

\begin{figure*}
      \centering
    \includegraphics[width = \textwidth]{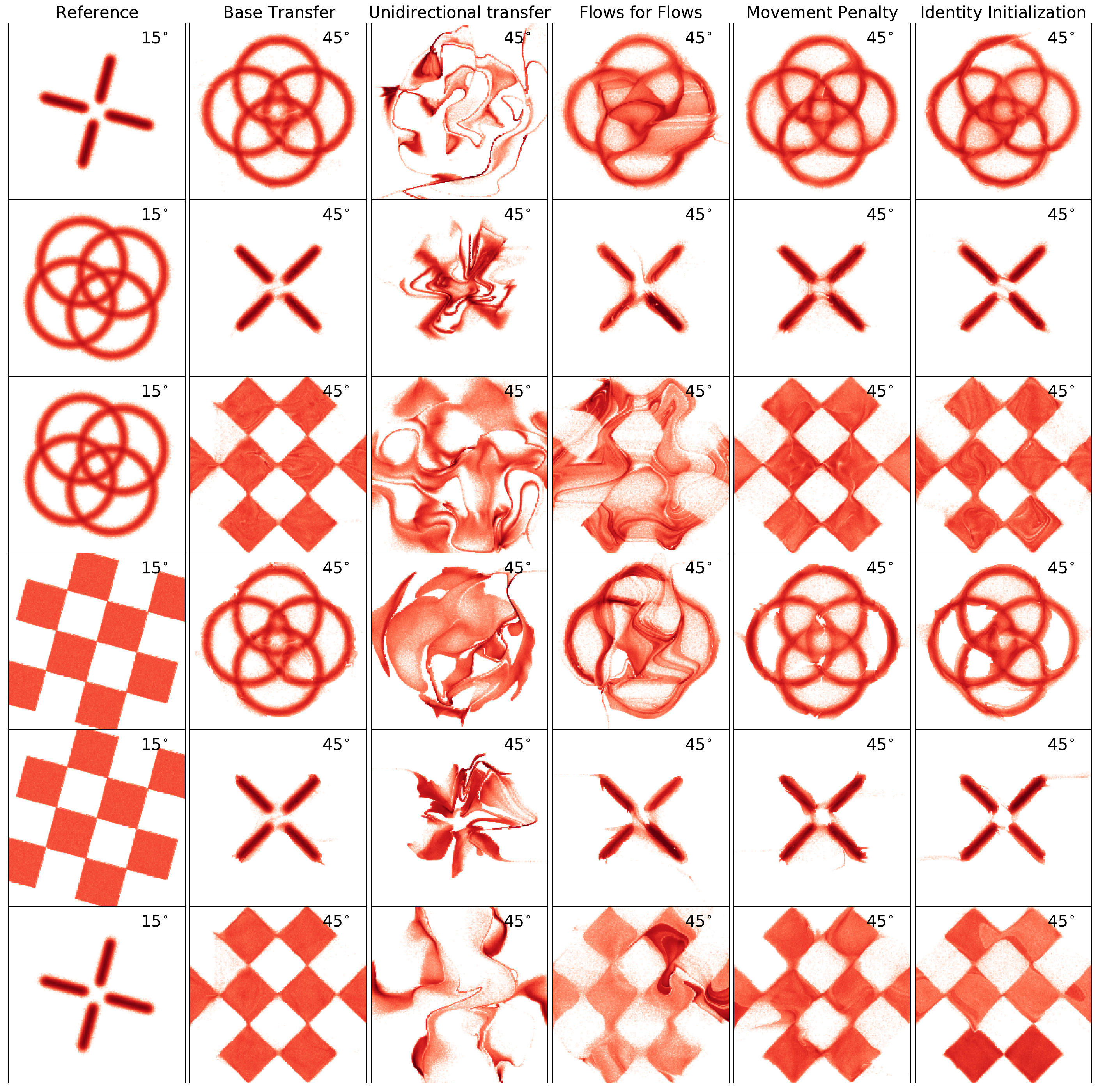}
    \caption{Transport tasks between various choices of identical reference and target toy distributions. The colorbar
has been set to scale logarithmically, which can emphasize out-of-distribution points. }
    \label{fig:summary_cond_transfer_15_00_45_00}
    
\end{figure*}

For a transport task between nonidentical toy distributions (shown in Fig. \ref{fig:fourcircles_star_cond_base_transfer_f4f_15_00_22_50}), the flows for flows method performs far more similarly to the base transfer method with respect to dispersing samples from the rings to the arms of the star. When considering just the distribution shapes (shown in Fig. \ref{fig:summary_cond_transfer_15_00_45_00}), the unidirectional transfer performs extremely poorly, the flows for flows method performs next best, the movement penalty and identity initialization methods perform well, and the base transfer method shows the cleanest final state. Again, given the more intuitive performance rankings for the scientific dataset, it is likely that the superior performance of the base transfer method is more a reflection of the contrived nature of the transport tasks between these toy distributions chosen.

\begin{figure*}
      \centering
    \includegraphics[width = \textwidth]{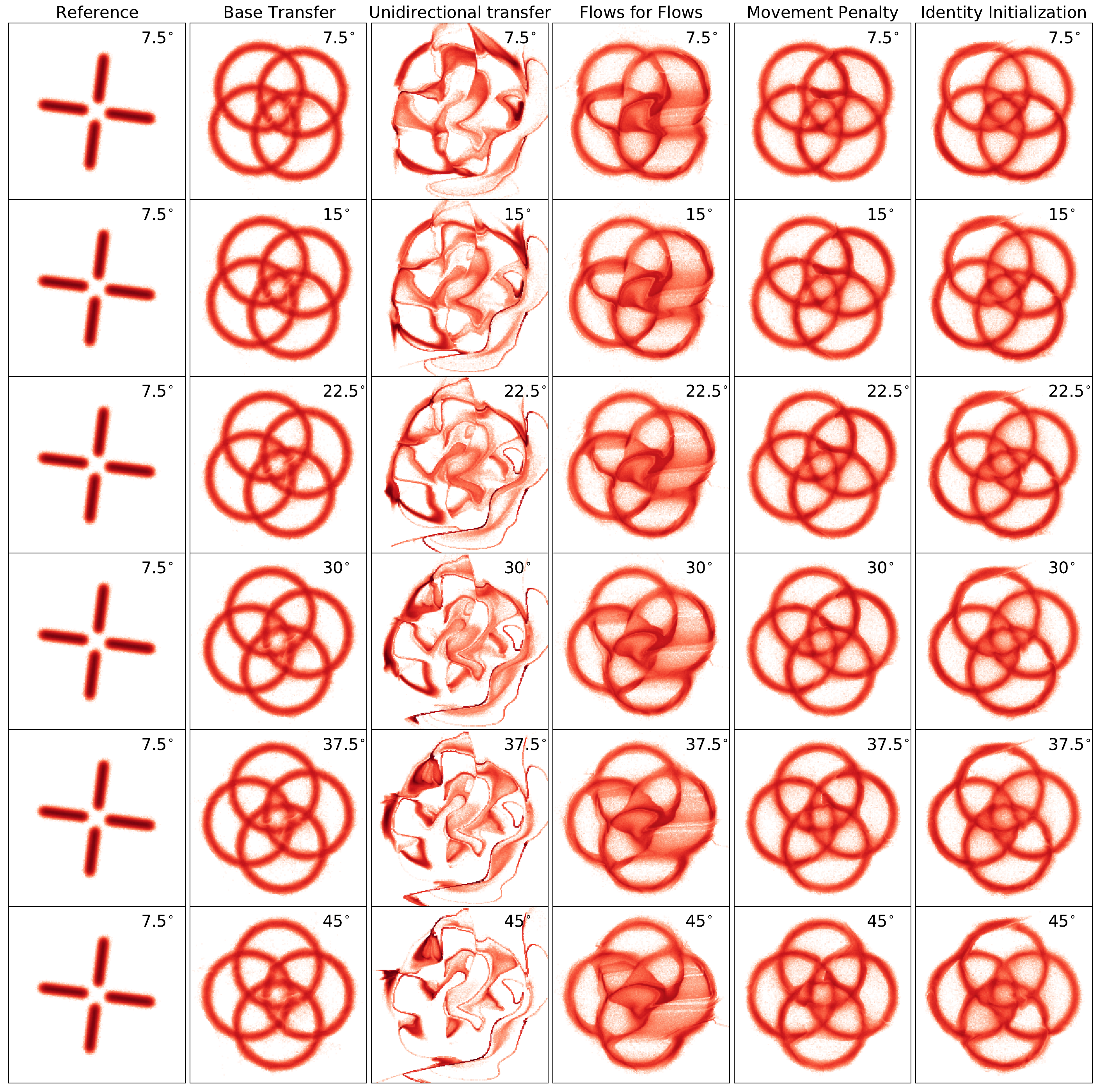}
    \caption{Transport tasks between the fourcircles and star distributions at a variety of conditioning angles. The colorbar
has been set to scale logarithmically, which can emphasize out-of-distribution points.}
    \label{fig:summary_cond_transfer}
    
\end{figure*}

In Fig. \ref{fig:summary_cond_transfer}, we show again a series of transport tasks between two nonidentical toy distributions, but for a selection of conditioning angles. The choice of angle does not appear to affect the performance of each individual transport method.

\end{document}